# Slow water in engineered nano-channels revealed by color-center-enabled sensing

Daniela Pagliero[1], Rohma Khan[1,2], Kapila Elkaduwe[2,3], Ankit Bhardwaj[4], Kang Xu[1], Abraham Wolcott[1,5], Gustavo López[2,6], Boya Radha[4,7,†], Nicolas Giovambattista[2,3,†], and Carlos A. Meriles[1,2,†]

Nanoscale confinement of molecules in a fluid can result in enhanced viscosity, local fluidic order, or collective motion. Confinement also affects ion transport and/or the rate and equilibrium concentration in a chemical reaction, all of which makes it the subject of broad interest. Studying these effects, however, is notoriously difficult, mainly due to the lack of experimental methods with the required sensitivity and spatial or time resolution. Here we leverage shallow nitrogen-vacancy (NV) centers in diamond to probe the dynamics of room-temperature water molecules entrapped within ~6-nm-tall channels formed between the diamond crystal and a suspended hexagonal boron nitride (hBN) flake. NV-enabled nuclear magnetic resonance measurements of confined water protons reveal a much reduced $H_2O$ self-diffusivity, orders of magnitude lower than in bulk water. We posit the slow dynamics stem from the accumulation of photogenerated carriers at the interface and trapped fluid, a notion we support with the help of molecular dynamics modeling. Our results provide feedback for theories describing interfacial water, and lay out a route for investigating other fluids under confinement.

Remarkable phenomena take place at the solid-liquid boundary, a region where traditional descriptions of the bulk properties of fluids and solids break down[1]. Confined liquids in particular tend to form partially ordered layers leading, e.g., to structural oscillatory solvation forces[2] and non-linear viscoelastic dynamics[3], important in areas spanning geophysics, tribology, catalysis, polymer science, and biology. Since the nanoscale characterization of interfacial and nanoconfined liquids is inherently complex, much effort has been devoted to engineering confining geometries amenable to controlled studies, often designed to reproduce nanofluidic transport processes governing the physics of living systems[4-6]. This interest also extends to the chemistry of nanoconfined solutions where fluctuations in molecule number play an important role in determining a reaction's outcome[7,8].

Several techniques have been used to examine the dynamics of interfacial molecules including electrochemical and optical methods, photoelectron and electron spectroscopies, ion scattering, field emission microscopy, and low-energy electron diffraction, to mention just some[9]. Out of these, however, only a few have the combined spatial and temporal sensitivity to probe ensembles locally. Techniques using atomic force or scanning tunneling microscopy can overcome the spatial resolution barrier encountered in X-ray diffraction, neutron scattering, or inductively detected magnetic resonance. Recent work has demonstrated the use of atomic force microscopy (AFM) to probe the rheology of liquids confined between the apex of the AFM tip and a solid substrate[10,11], which has proven useful, e.g., to examine theoretical nanoscale models of molecular friction[12,13]. Unfortunately, AFM requires direct access to the sample and ranks as an invasive technique, ill-adapted for sealed fluidic systems or complex nanoscale environments. Further, the temporal resolution of AFM is poor, thereby limiting the experimenter's ability to probe dynamical processes.

In this study, we demonstrate ¹H nuclear magnetic resonance (NMR) of water molecules confined to 185-nm-wide, 5.6-nm-tall channels formed between the surface of a diamond crystal and a suspended hexagonal boron nitride (hBN) flake. To circumvent the sensitivity limitations inherent to inductively detected NMR, we resort to a small ensemble of near-surface nitrogen-vacancy (NV) centers, here acting as optically-detected magnetic sensors embedded in the diamond substrate. Sensitive to water mobility, our ¹H NMR experiments reveal $H_2O$ self-diffusivity orders of magnitude below that found in bulk water. Molecular dynamics (MD) simulations indicate confinement alone is insufficient to account for the observed dynamics, and supports instead the notion of an electric-field-induced process resulting from the accumulation of photo-generated excess carriers both on the diamond surface and in the confined fluid.

[1]Department of Physics, CUNY–The City College of New York, New York, NY 10031, USA. [2]CUNY-Graduate Center, New York, NY 10016, USA. [3]Department of Physics, CUNY–Brooklyn College of the City University of New York, Brooklyn, NY, 11210, USA. [4]Department of Physics and Astronomy, The University of Manchester, Manchester, UK. [5]National Graphene Institute, The University of Manchester, Manchester, UK. [6]Department of Chemistry, San José State University, San José, CA 95192, USA. [7]Department of Chemistry, CUNY–Lehman College, Bronx, NY 10468, USA. †E-mail(s): radha.boya@manchester.ac.uk, NGiovambattista@brooklyn.cuny.edu, cmeriles@ccny.cuny.edu.

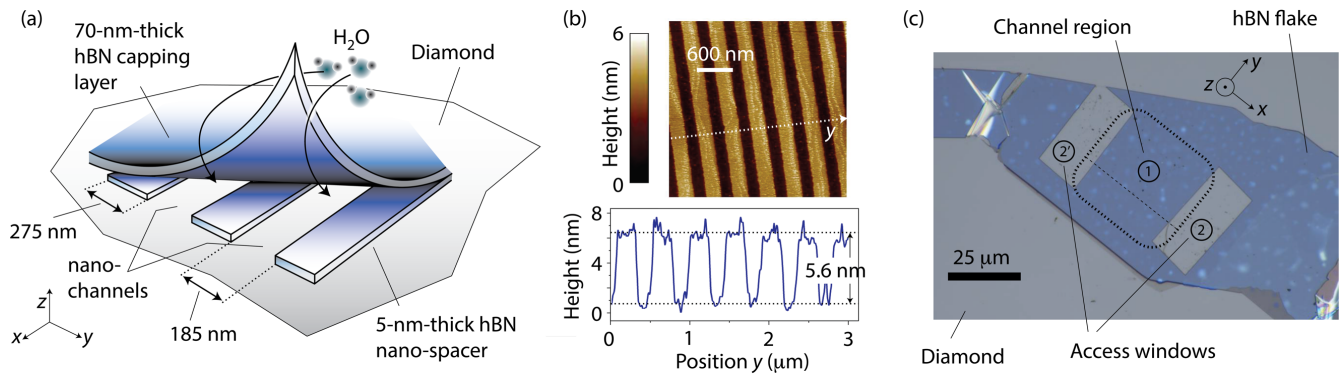

**Figure 1 | Nanoscale confinement of water molecules.** (a) Schematics of the nanochannel structure. We overlay a ~70-nm-thick hBN flake on a set of 5.6-nm-tall hBN spacers previously engineered on the surface of a [100] diamond via electron beam lithography. The crystal sits within a sealed chamber in a microfluidic device (not shown), hence allowing us to control the sample environment. (b) Atomic force microscopy image of a channel section prior to adding the capping hBN layer; the bottom insert is a cross-sectional plot across the dashed line. (c) Scanning electron microscopy image of the complete structure. The dashed square (here denoted as area ①) indicates the nanochannel section; the dashed line runs parallel to the channels and water intake takes place through two rectangular side openings (areas ②).

## $^1$H NMR of confined water

Figure 1a shows a schematic of the nanochannel structure: We use electron beam lithography to pattern a set of 5.6-nm-tall strips of hBN (see also Fig. 1b), which we then cap with a hBN flake sufficiently thick (70 nm) to avoid sagging[14]; windows on either end of the channels serve as inlet/exit ports (Fig. 1c). To gain control over the environment, we enclose the structure in a microfluidic chip, and fill in the nanochannels by prolonged exposure to saturated water vapor (Supplementary Material (SM), Section I).

We combine chemical vapor deposition overgrowth, $^{15}$N implantation, and thermal annealing to cap the starting seed diamond with a $^{13}$C-depleted layer hosting ~8-nm-deep NV centers (Fig. 2a). Formed by a substitutional nitrogen adjacent to a vacancy, these paramagnetic point defects have emerged as versatile magnetic sensors whose spin state can be initialized, manipulated, and read out through combined laser light and microwave (MW) excitation[15] (Fig. 2b). For the present experiments, we apply a magnetic field in the 38–45 mT range along one of the four possible crystalline axes, and tune the MW to the $|m_{NV} = 0\rangle \leftrightarrow |m_{NV} = -1\rangle$ NV spin transition[16]. For illustration, the plot in Fig. 2b shows an optically detected magnetic resonance (ODMR) NV spectrum, often manifesting as an asymmetric doublet due to nuclear spin pumping of the [15]N host (SM, Section I).

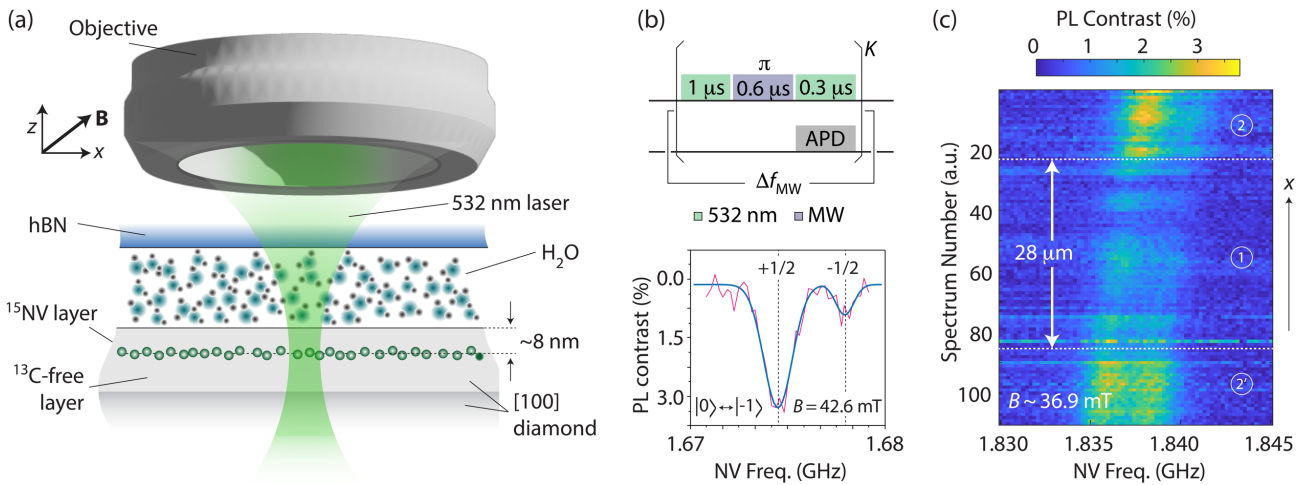

**Figure 2 | Nanoscale sensing via shallow NV centers.** (a) We implement nanoscale $^1$H NMR with the help of an 8-nm-deep layer of engineered $^{15}$NV centers serving as magnetic sensors. (b) (Top) Pulsed ODMR protocol; we average the NV PL after $K$ repeats of a MW π-pulse, whose frequency we vary in steps $\Delta f_{MW}$ to reconstruct the NV spin resonance spectrum. (Bottom) ODMR of the $|m_{NV} = 0\rangle \leftrightarrow |m_{NV} = -1\rangle$ $^{15}$NV transition. The vertical dashed lines mark the frequencies expected for the PL dips corresponding to the $m_N = \pm 1/2$ projections of the hyperfine coupled $^{15}$N host, and the solid trace is a Gaussian fit. Careful alignment of the externally applied magnetic field **B** along one of the crystalline NV axes — at 54.7 degrees relative to the crystal normal $z$ — leads to near-full $^{15}$N spin polarization. (c) NV ODMR along a line parallel to the channels (area ①) and connecting the two access windows (areas ②, see dashed line in Fig. 1c). Notably, hBN partially quenches the NV spin contrast. APD: Avalanche photo detector. MW: Microwave. Freq.: Frequency.

NV sensing in the nanochannel area is especially challenging because the hBN structure has a detrimental impact on the NV spin and optical properties. As an illustration, Fig. 2c shows the NV ODMR spectrum at different positions along a line running parallel to the channels from one access window to the other (see dashed line in Fig. 1c): Besides the slight frequency shift and changing amplitude ratio in the resonance doublet (a mere consequence of magnetic field heterogeneity across the structure), we find an overall reduction of the photoluminescence (PL) contrast. We interpret this behavior as a likely consequence of surface-induced NV charge dynamics, a subject we return to later (see also SM, Sections I and II).

To measure the [1]H NMR spectrum from water molecules, we implement a dynamical decoupling sensing protocol in the form of a MW train of inversion pulses separated by a time interval $\tau$ (Fig. 3a). For sufficiently long MW trains — the case in our experiments — NVs become selectively sensitive to magnetic fields oscillating within a small frequency window centered at $1/2\tau$[17]. Correspondingly, a signature NMR signal develops as one varies the inter-pulse spacing to match half the [1]H spin precession period at the externally applied field[18,19]. Figure 3a shows representative examples before and after exposing the channels to saturated water vapor: Inspecting a site in the nanochannel area, we observe a clear PL dip at the [1]H Larmor frequency, hence revealing the presence of entrapped water. Slow evaporation gradually empties the channels over a time span of 2 to 3 weeks (SM, Section I) implying the weak NMR signal we observe before the fill (left panel in Fig. 3a) originates from residual water entrapped during prior runs.

**Water diffusion under nanochannel confinement**

The ability to probe unpolarized water protons via NV centers — to our knowledge, shown here for the first time — is intriguing because H$_2$O molecules are expected to diffuse too fast to encode an observable signal. For example, using the self-diffusion coefficient of bulk water $D^{(b)} = 2.3 \times 10^{-9}$ m$^2$ s$^{-1}$ as a reference[20], one expects water molecules to leave the NV sensing range $d$ after a time $T_D^{(b)} \sim d^2/6D^{(b)}$; since $d$ must be comparable to the NV depth[19], we find $T_D^{(b)} \sim 5$ ns, orders of magnitude shorter than the μs-duration of the MW pulse trains required for nuclear spin sensing.

To gauge the diffusion dynamics of water more accurately, we implement a spin correlation measurement whose output reflects on the [1]H spin coherence after a variable evolution time $\tilde{\tau}$[21]. Molecular diffusion during this interval amounts to a phase disruption in the nuclear spin set being sensed, and thus leads to a reduction in the observed NV signal amplitude. Because decoherence from internuclear dipolar couplings is weak for a fluid, molecular self-diffusion governs the overall decay whose characteristic time $T_D$ can be seen to represent the system's diffusion time[22,23].

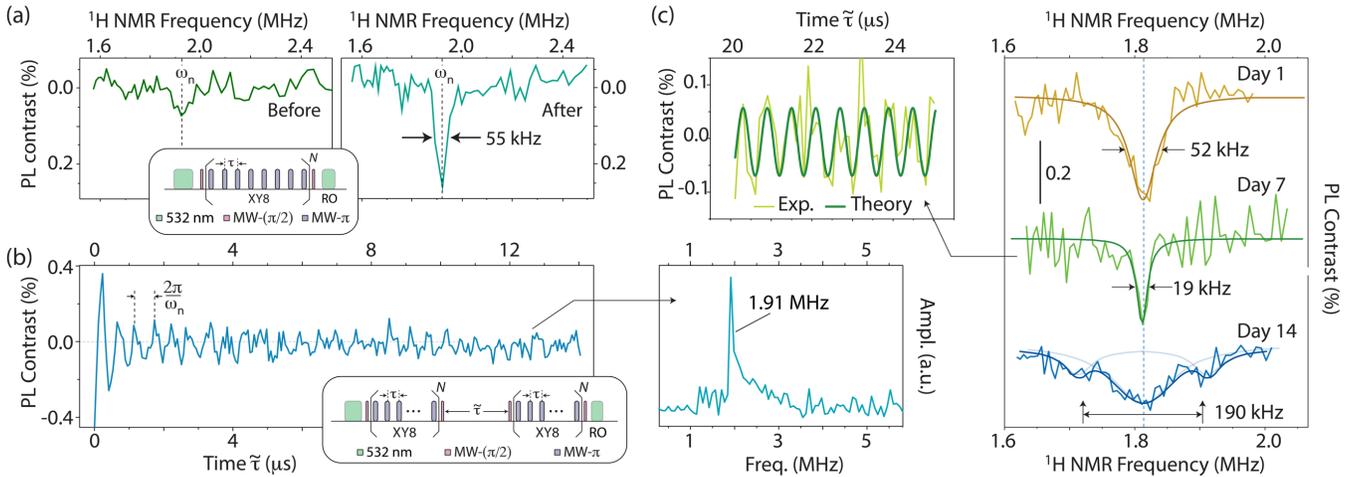

**Figure 3 | [1]H NMR of confined water.** (a) [1]H NMR spectra before and after filling the channels with water (left- and right-hand side plots, respectively). The vertical dashed line indicates the expected [1]H Larmor frequency $\omega_n$. The small dip prior to water exposure is likely a leftover from prior fillings. We use $N = 10$ XY8 decoupling units, and apply $4 \times 10^6$ repeats for each inter-pulse interval $\tau$. (b) [1]H spin correlation signal after channel filling and corresponding Fourier transform (right-hand side insert). The conditions are those of (a) except that $N = 5$. (c) (Right panel) NV-detected [1]H NMR spectra of confined water at different times (1, 7, 14 days) after filling the channels. Smooth solid traces are Lorentzian fits, and the vertical dashed line indicates the expected [1]H Larmor frequency $\omega_n$. (Left panel) [1]H correlation signal one week after channel filling (green trace in the main panel); the solid trace is sinusoidal at the [1]H Larmor frequency. We observe a nuclear spin coherence exceeding 25 μs of free evolution with no apparent decay. RO: Readout. Ampl.: Signal amplitude.

Figure 3b shows the results: We observe a distinctive oscillatory response centered at the $^1$H Larmor frequency whose envelope features a rapid initial decay followed by a long-lived tail, a non-exponential dependence inherent to the single-sided geometry of NV center sensing[23]. A detailed analysis shows a bi-modal envelope is necessary to describe the measured time dependence with fast and slow time constants $T_D^{(f)} = 0.5$ μs and $T_D^{(s)} = 30$ μs (SM, Section II). We derive, therefore, diffusion constants $D^{(f)} \approx d^2/6T_D^{(f)} \sim 2.1 \times 10^{-11}$ m$^2$ s$^{-1}$ and $D^{(s)} \sim 0.4 \times 10^{-12}$ m$^2$ s$^{-1}$, orders of magnitude lower than in bulk water.

We caution that there is substantial variability in the evolution of the spectra after channel filling, which suggests complex dynamics, possibly impacted by uneven evaporation over time. For example, the right panel in Fig. 3c showcases an instance where the NMR linewidth (52 kHz at the time of the channel refill, yellow trace) shrinks to less than half the original value after a week-long interval (green trace). Correlation measurements at that time revealed long-lived $^1$H coherences extending over tens of microseconds with minimal decay; from the trace in the left panel of Fig. 3c, we derive $T_D \geq 100$ μs, and correspondingly, $D \leq 0.1 \times 10^{-12}$ m$^2$ s$^{-1}$.

Subsequent observations after a two-week span showed a much broader NV-NMR spectrum whose reduced amplitude prevented correlation measurements (blue trace in Fig. 3c). On the other hand, new spectral features in the form of weak satellites flanking the main dip suggest this broadening may not necessarily stem from faster molecular diffusion, but rather be the result of other interactions at play. In particular, the magnitude of these splittings (largely exceeding the sub-1-kHz range reported for dipolarly-coupled water protons on a silica substrate[24]) hint at couplings with electronic spins, known to reach hundreds of kHz for water protons in the solvation layer around metal ions[25].

To rationalize all these observations, we posit that water diffusivity slows down through confinement-aided buildup of interfacial charge created by the photo-injection of carriers from the surrounding walls. Charge injection from shallow color centers — already seen in diamond[26,27] and likely present in hBN — is a consequence of the optical excitation inherent to our experiments, affecting the charge state of surface traps and other coexisting point defects. Shallow NV centers in particular undergo cyclic photo-ionization and recombination under green illumination[28], a process recently exploited to demonstrate solvation of carriers into water in contact with the diamond surface[27].

While the sign of the carriers exiting the crystal is presently unclear, the negative surface affinity of chemically oxidized diamond[26,29] along with the upward band bending characteristic of NV-engineered surfaces[30,31] suggest that positive charge accumulates on the diamond side of the interface while electrons diffuse into the liquid[32,33]. We hypothesize that this process leads to (i) the formation of an electric double layer and, ultimately, to a higher degree of self-organization that slows down the H$_2$O dynamics at the interface. In principle, it is also possible that (ii) uneven evaporation within the channels results in sections of the confining surfaces being covered only by a few layers of water in contact with vapor domains, which could also help slow down water motion. Our computer simulations support (i) but not (ii).

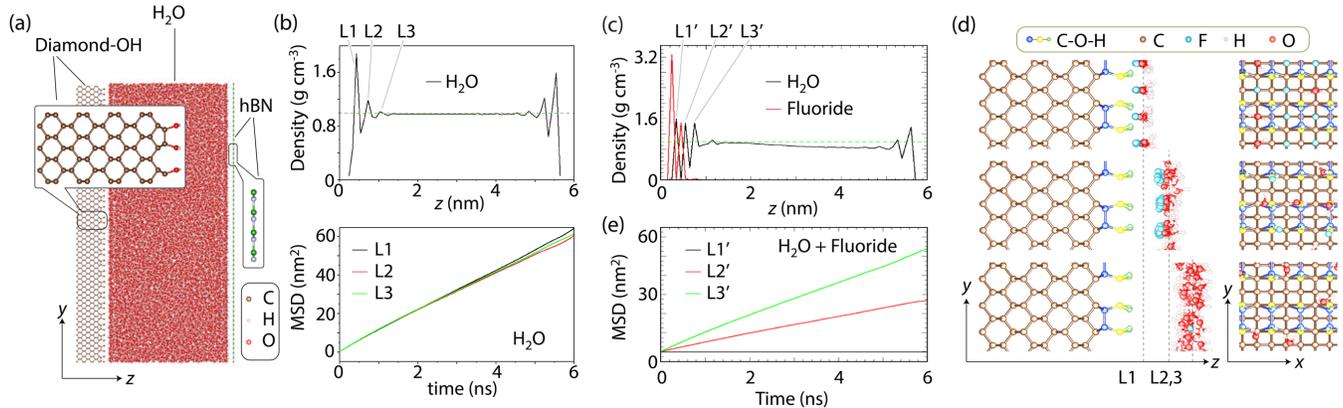

**Figure 4 | Molecular dynamics simulations of confined water with and without added charges.** (a) Snapshot from MD simulations of a system comprising $N_w = 30442$ water molecules confined by an hBN monolayer and a hydroxylated diamond at temperature $T = 300$ K. (b) (Top) Water density profile as a function of the distance $z$ from the diamond surface for the system shown in (a). (Bottom) Time dependence of the mean-square displacement (MSD) for water molecules initially within layers L1 ($z < 0.55$ nm), L2 ($0.55 < z < 0.95$ nm), and L3 ($0.95 < z < 01.25$ nm) next to the diamond crystal. (c) Calculated water and fluoride density profiles as a function of the distance $z$ from the diamond surface after replacing $N_q = 2704$ water molecules in (a) by fluoride anions; we attain charge neutrality by distributing positive charges in the diamond (SM, Section III). The green dashed line indicates the bulk density of water under ambient conditions. (d) Side and top views of layers L1' through L3' defined in (c). (e) MSD as a function of time for water molecules initially located within layers L1' ($z < 0.45$ nm), L2' ($0.45 < z < 0.65$ nm), L3' ($0.65 < z < 0.85$ nm). Molecules within L2 and L3 (respectively, red and green traces) diffuse away while molecules in L1 do not (black trace).

**Molecular dynamics simulation of confined water**

To establish a starting reference, we first study the dynamics of pure water confined to a geometry emulating our experimental conditions (Fig. 4a). Given the compositional heterogeneity of chemically oxidized diamond surfaces[34], we consider four different terminations[35] (H, OH, O, and H-OH, see SM, Section III). The top panel in Fig. 4b displays the calculated water density profile as a function of the relative position $z$ between the confining walls for the hydroxylated diamond surface. Similar to prior experimental[36-38] and MD simulations[39], proximity to the walls leads to the formation of discrete molecular layers — L1 through L3 — but the order extends only over distances of ~1 nm, smaller than the channel height. Further, the bottom plot in Fig. 4b shows the mean-square displacements (MSD) as a function of time for water molecules initially in each layer: In all cases, we find the MSD exceeds 60 nm$^2$ within 6 ns, thus indicating water molecules leave the sensing volume with rates characterized by bulk-like diffusion coefficients. We find similar results when we consider 1-nm-thick water films produced by partial evaporation (SM, Section III). It follows that MD simulations of nanoconfined (pure) water are *incompatible* with the observed NV-NMR signal buildup.

Modeling the impact of solvated electrons on the water dynamics is challenging given the transient formation of water structures that can only be described via rigorous quantum mechanical descriptions[40], conflicting with the mesoscale size of our system. For simplicity, here we ignore chemical transformations and use instead classical MD simulations to recreate the effect of electrostatic forces created by injected charges. To this end, we consider a model system where water and fluoride anions ($F^-$) coexist in the nanochannels (SM, Section III).

Figure 4c shows the calculated equilibrium density profiles of water and $F^-$ ions at $T = 300$ K for hydroxylated diamond. As in the pure water case, H$_2$O molecules organize to form layers parallel to the solid walls. The diamond surface features regularly spaced atomic scale pockets that can be filled by either water molecules or $F^-$ ions. We find these pockets are preferentially occupied by the ions, followed by a first monolayer of water molecules (Fig. 4d). To assess the impact of these charges on the water dynamics, we calculate the MSD of the water molecules that are initially located in layers L1', L2' and L3'. As shown in Fig. 4e, the MSD of first-layer water is negligible within the 6-ns span of the simulation suggesting that the molecules in this layer are effectively immobile; diffusion is gradually faster for successive water layers, though the overall dynamics are much slower than observed in the absence of ions. Similar results are obtained with diamond surfaces terminated with H, O, and H+OH groups (see SM, Section III). These findings may relate to preceding MD simulations of water in the presence of externally applied electric fields, found to stabilize molecular motion to ultimately induce a phase transition into an ice-like structure[41].

Interestingly, proximity of nuclear spins to surface charges may potentially lead to hyperfine interactions strong enough to broaden the observed NMR spectrum, hence providing a rationale for spectral broadening even in the presence of reduced water diffusivity (possibly the case in the bottom trace of Fig. 3c). Discrete satellites in the spectrum suggest the presence of preferential, long-lived water structures, though its physical nature is presently unclear. Along the same lines, a uniform distribution of hyperfine couplings — corresponding to multiple coexisting water structures coupling to carriers to a varying degree — can help explain the asymmetric shape in the spectrum extracted from the $^1$H correlation signal in Fig. 3b (SM, Section II).

**Conclusions**

In summary, we demonstrated proton NMR of water molecules confined to nanochannels engineered on a diamond surface. Long-lasting nuclear spin coherences point to slow molecular self-diffusion, up to four orders of magnitude smaller than predicted by MD simulations that only consider spatial confinement. Space charge fields emerging from carrier photogeneration and injection into the fluid may underlie these findings. While observations away from the channel area yield negligible proton NMR signal (SM, Section I), additional work will be needed to clarify the role of the confining structure and whether similar dynamics can be induced in water layers adsorbed on the diamond surface.

Future experiments include extensions to temperatures down to the freezing point or in the presence of external electric fields. Structures where the channel height diminishes with the amount of entrapped water[42] could also prove useful to controllably reach the limit of individual atomic layers, where water is expected to show a rich phase behavior[43]. By the same token, simulations combining quantum ab-initio MD, Monte Carlo, and machine learning could help implement more rigorous models without the tradeoff imposed by the mesoscale size of the system[43]. Our results may prove relevant to the implementation of biomimetic computations on aqueous electrolytic chips[44], the development of novel nuclear spin hyper-polarization techniques based on spin polarized carriers[27], and the understanding of water dynamics proximal to large biomolecules[45].

**Data availability**

The data that support the findings of this study are available from the corresponding author upon reasonable request.

**Code availability**

All source codes for data analysis and numerical modeling used in this study are available from the corresponding author upon reasonable request.

**Acknowledgments**

We acknowledge helpful discussions with Alexander Wood, Alexander Pines, Joerg Wrachtrup, and Durga Dasari. K.X., K.E.R.K.W., N.G., and C.A.M. acknowledge support by the National Science Foundation through grant NSF-2223461; similarly, D.P. acknowledges support from grant NSF-2203904. The work of R.K. and A.W. was supported by CREST-IDEALS, NSF-2112550. B.R. and A.B. acknowledge funding from the European Union's H2020 Framework Programme/ERC Starting Grant 852674 AngstroCAP, Royal Society University Research Fellowship URF\R\231008, Philip Leverhulme Prize PLP-2021-262, EPSRC new horizons grant EP/X019225/1, and EPSRC strategic equipment grant EP/W006502/1. All authors acknowledge access to the facilities and research infrastructure of the NSF CREST IDEALS, grant number NSF-2112550.

**Author contributions**

D.P., R.B., N.G., and C.A.M. conceived the experiments. D.P., R.K., and K.X. conducted the experiments with assistance from A.W.; K.E.R.K.W. led the molecular dynamics modeling with assistance from G.L., and N.G.; A.B. fabricated the nanochannel sample under the supervision of R.B. All authors analyzed the data; C.A.M. wrote the manuscript with input from all authors.

**Competing interests**

The authors declare no competing interests.

**Correspondence**

Correspondence and requests for materials should be addressed to C.A.M.

# Supplementary Material for

# Slow water in engineered nano-channels revealed by color-center-enabled sensing

Daniela Pagliero[1], Rohma Khan[1,2], Kapila Elkaduwe[2,3], Ankit Bhardwaj[4,5], Kang Xu[1], Abraham Wolcott[1,6], Gustavo López[2,7], Boya Radha[4,5,†], Nicolas Giovambattista[2,3,†], and Carlos A. Meriles[1,2,†]

[1]*Department of Physics, CUNY–The City College of New York, New York, NY 10031, USA.*
[2]*CUNY-Graduate Center, New York, NY 10016, USA.*
[3]*Department of Physics, CUNY–Brooklyn College of the City University of New York, Brooklyn, NY, 11210, USA.*
[4]*Department of Physics and Astronomy, The University of Manchester, Manchester, UK.*
[5]*National Graphene Institute, The University of Manchester, Manchester, UK.*
[6]*Department of Chemistry, San José State University, San José, CA 95192, USA.*
[7]*Department of Chemistry, CUNY–Lehman College, Bronx, NY 10468, USA.*

[†]Corresponding authors: radha.boya@manchester.ac.uk, NGiovambattista@brooklyn.cuny.edu, cmeriles@ccny.cuny.edu.

# I. Materials and experimental methods

## I.1 *Nanochannel fabrication*

Our nanometer-slit device comprises two hexagonal boron nitride (hBN) flakes, prepared by mechanical exfoliation from bulk, high-purity hBN[1]. We first exfoliate a 5.6 nm thick layer and transfer it to a silicon wafer featuring a 90 nm layer of $SiO_2$; we subsequently resort to electron beam lithography (Zeiss SmartSEM + Raith ELPHY Quantum NNANOSUITE V6.0) and dry etching (Plasmalab 100 ICP-65, Oxford) to produce a two-dimensional (2D) striped array, as shown in Fig. 1 of the main text. Collectively, the hBN strips — each approximately 275 nm wide, separated from nearest neighbors by 185 nm — function as spacers defining the channels height[2]. To cap the array, we transfer a 70-nm-thick mechanically-exfoliated hBN flake to another $SiO_2$ substrate; this thickness provides sufficient mechanical rigidity to prevent sagging of the slits in air or swelling in water. After each transfer, we bake the device on a hotplate at 170°C for 30 minutes to remove any absorbates and provide good adhesion; this includes the transfer of the of the top-spacer stack to its final diamond substrate, which we carry out via adhesion to poly-methyl-methacrylate (PMMA). Once on the diamond, we use electron beam lithography and dry etching to open two windows on either end of the spacers, hence allowing $H_2O$ molecules to access the channels upon exposure to water vapor. Prior to magnetic resonance measurements, we use an optical microscope (Nikon Eclipse LV100ND) to observe hBN flakes after exfoliation and for device inspection at all stages. We also employed atomic force microscopy (Dimension Fastscan with Bruker's Scanasyst) to probe the height profile of the hBN spacers. We summarize the fabrication process in Fig. S1.

## I.2 *Operating geometry and channel filling protocol*

The diamond crystal sits within the main chamber of a 3D printed microfluidic cell sealed with transparent windows on both sides to allow for optical measurements (Fig. S2). A through channel and two side ports connect the chamber to the outside, hence allowing for a controlled environment. A double, 0.5-mm-diameter loop of thin (44 gauge) coated copper wire glued to one of the windows serves as the MW antenna.

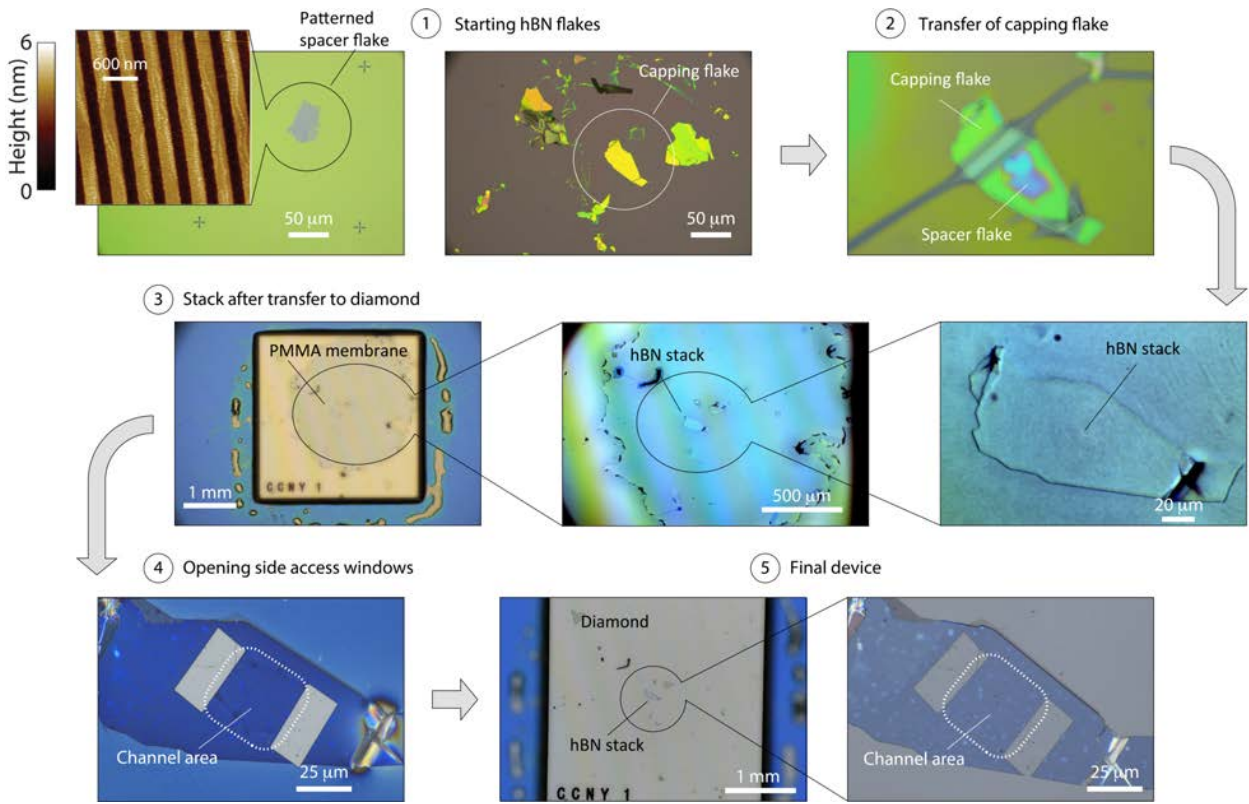

**Figure S1 | Nanochannel fabrication.** Starting with hBN flakes on separate $SiO_2$ substrates, we first use electron beam lithography to produce a striped, 5.6-nm-tall spacer pattern, which we then cap with a 70-nm-thick flake for upper channel confinement (steps ① and ②, the dark diagonal shadow in ② is a PMMA membrane holding the top flake). We then use PMMA to transfer the stack into diamond (step ③), where we open two side windows via electron beam lithography and dry etching (step ④). The last two images show the end structure after PMMA removal (step ⑤).

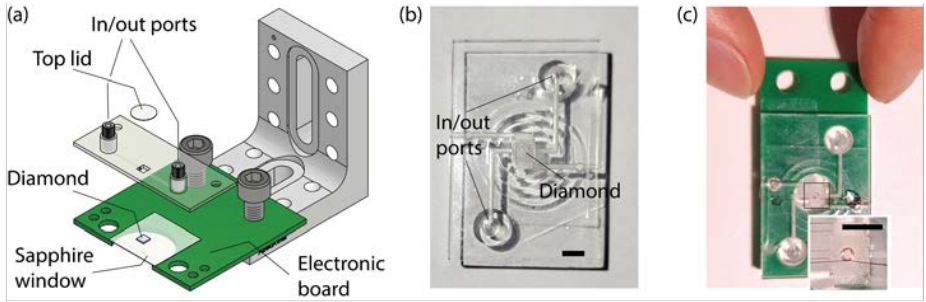

**Figure S2 | Controlling the nanochannel environment.** (a) Schematics of the sample mount including the microfluidics chip and antenna board. (b) Image of the microfluidic chip. (c) Photograph of the final device. The insert is a zoomed-out image of the squared section. The scale bar in (b) and (c) is 2 mm.

To water-fill the channels, we first flush MilliQ H$_2$O through the ports to flood the sample chamber, a wetting step lasting for about an hour. We then flow argon gas — previously humidified via bubbling in MilliQ water within a separate container — through the chip's inlet port and into the sample chamber. This ~12-hour-long step removes bulk water from the chamber while maintaining a humid atmosphere. No gas flow takes place during measurements.

**I.3** *Experimental setup*

We use a home-built confocal microscope with a 532 nm laser and an acousto-optic modulator (AOM) to control laser pulsing with time precision down to 10 ns. Sample illumination and photon collection rely on a 0.9-numerical-aperture Olympus objective with 100× magnification; the laser power reaching the NVs is typically 3 mW. We use a galvo mirror system for XY scans and beam positioning; we achieve Z control with a piezoelectric stage. A white light source steered into the same objective allows us to alternatively obtain optical images, which we use to ensure correct positioning of the laser beam on the channel area. We detect the NV fluorescence via a 25-µm-fiber-coupled avalanche-photo-detector (APD-Excelitas) after a dichroic mirror and a long-pass filter; we use a SpinCore PulseBlaster to gate the APD and the AOM. To manipulate the NV spin, we apply pulsed microwave (MW) excitation from a signal generator (Rohde & Schwarz). We feed the I and Q channels from an arbitrary wavefunction generator (AWG-SE5082 from Tabor electronics) into an IQ mixer (ADL5375, 400 MHz to 6 GHz Broadband Quadrature Modulator), which allows us to attain MW phase control. Sample MW excitation rely on an amplifier (Minicircuits ZHL-25W-272+) and an on-chip, 50-Ohm-terminated antenna (see above). We use a permanent magnet to create a magnetic field of ~40 mT along a direction forming an angle of 54.6 deg. relative to the surface normal and parallel to one of the NV crystallographic axes. All experiments take place under ambient conditions.

**I.4** *NV spin characterization and control*

We use a 2×2×0.5 mm$^3$, [100] electronic grade diamond crystal from E6, subsequently processed by QuantumDiamond. The crystal was overgrown with a $^{12}$C-enriched (99.99%), 1-µm-thick layer and then polished to bring down the average surface roughness to less than 1 nm. $^{15}$N-implantion (1×10$^{12}$ cm$^{-2}$, 3 keV) followed by thermal annealing (850 $^0$C) resulted in an 8-nm-deep NV layer with approximate concentration of 10$^2$ centers per µm$^2$. Prior to fabricating the nanochannel structure, we followed standard acid boiling protocols for chemical surface oxygenation; we limit subsequent surface cleanings to acetone immersion so as to preserve the device integrity. Standard PL imaging of the device in the channel area shows exposure to water — both in liquid and vapor forms — has a significant impact on the system brightness. In particular, we find the NV ensemble fluorescence changes after channel filling to gradually become dimmer, the state during measurements (Fig. S3a). These changes are largely reproducible and hence suggest charge state conversion processes with efficiency dictated by local heterogeneities in defect concentration and/or surface termination. We warn, however, there is some residual variability in the PL images we obtain in any given water-exposure-and-measurement cycle, implying it is challenging to ensure we examine the exact same (diffraction-limited) sites in subsequent experimental rounds (see below).

We use MW tuned to the $|m_S = 0\rangle \leftrightarrow |m_S = -1\rangle$ transition (see Fig. S3b for a schematic energy diagram). Note that in the limit of alignment between the applied magnetic field and the chosen NV crystallographic axis, optical illumination leads to $^{15}$N spin polarization, thus resulting in a narrower transition (Fig. 2b in the main text). With the exception of pulsed ODMR measurements, however, we make use of short pulses (typically, ~70 ns, Fig. S3c) for broadest spectral excitation, hence rendering $^{15}$N spin pumping convenient but unnecessary. We attain ~4% spin contrast during NV optical readout (see Rabi amplitudes in Fig. S3c), and measure characteristic Hahn-echo (dynamical decoupling) coherence times reaching 4.5 µs (19.8 µs), typical for shallow NVs at the present concentration[3] (Figs. S3d and S3e, respectively).

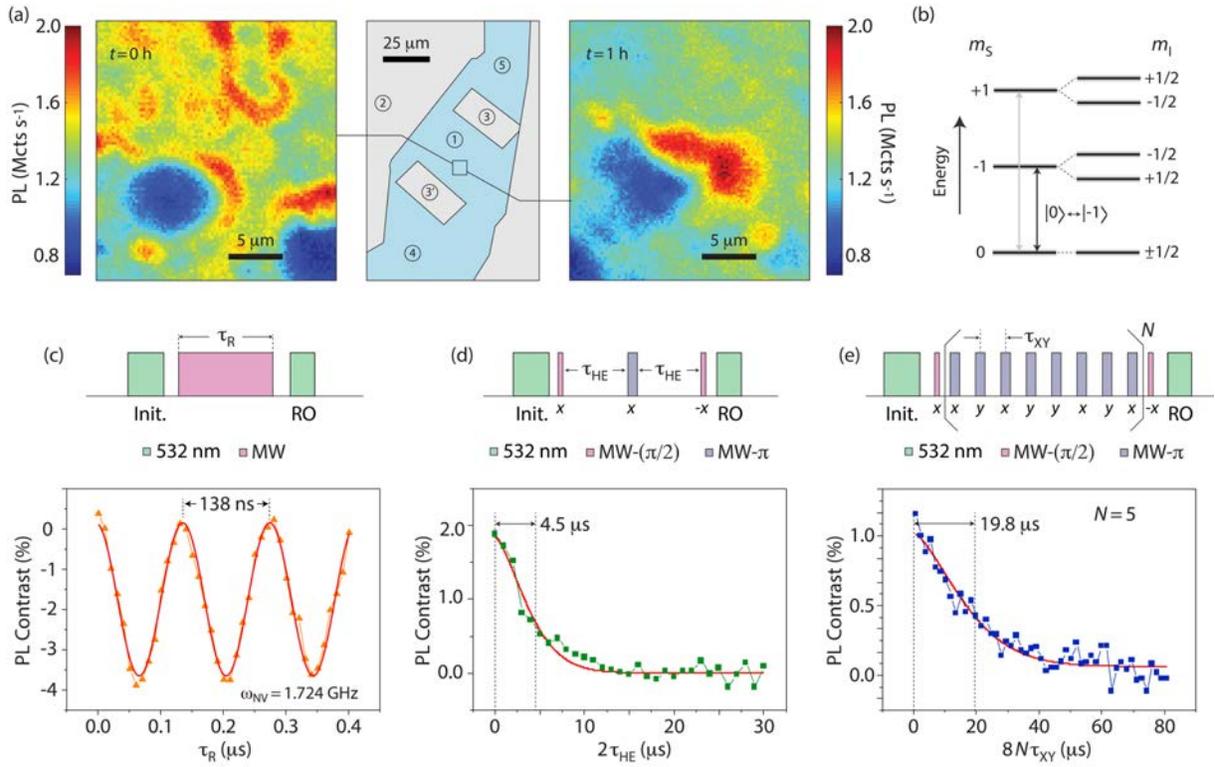

**Figure S3 | NV characterization.** (a) NV fluorescence imaging under green excitation of a region in the channel area immediately after exposure to liquid water and humidified argon (left) and an hour later (right). Significant changes in the PL brightness suggest a modification of the NV charge state. (b) Simplified energy diagram for the NV spin in the ground state orbital assuming a magnetic field $B$ aligned with the NV axis. Throughout our experiment, we tune the MW to the $|0\rangle \leftrightarrow |-1\rangle$ transition at the applied $B$; note that hyperfine coupling with the host $^{15}$N — a spin-1/2 nucleus — leads to splitting of the ODMR dip (Fig. 2b in the main text). (c through e) Characteristic NV response to Rabi, Hahn-echo, and XY8-$N$ protocols. In all cases, the solid red trace indicates a fit in the form of a damped sinusoidal (Rabi measurements) or a stretched exponential (Hahn-echo and XY8-5). In all protocols, the typical readout (RO) and initialization (Init.) times are 0.75 and 10 µs, respectively; the 532-nm laser intensity is 1.5 mW in all pulsed experiments.

## I.5 *NV-detected $^1$H magnetic resonance*

We implement $^1$H nuclear magnetic resonance (NMR) sensing with the help of XY8-$N$ dynamical decoupling protocols, selectively sensitive to AC magnetic fields whose frequencies match twice the time $\tau_{XY}$ separating consecutive pulses in the train[4]; an NMR spectrum emerges as one records the NV spin signal for varying $\tau_{XY}$ in the vicinity of the $^1$H Larmor frequency, $\omega_n$, at the applied magnetic field (typically in the range 38–45 mT). Correlation measurements use the same nuclear-spin-noise-sensing principle[5,6], except that this time it is the interval $\tilde{\tau}$ separating the encoding and decoding dynamical decoupling blocks that varies in subsequent runs; the end result is the equivalent of a "free induction decay" at the frequency defined by $\tau_{XY}$.

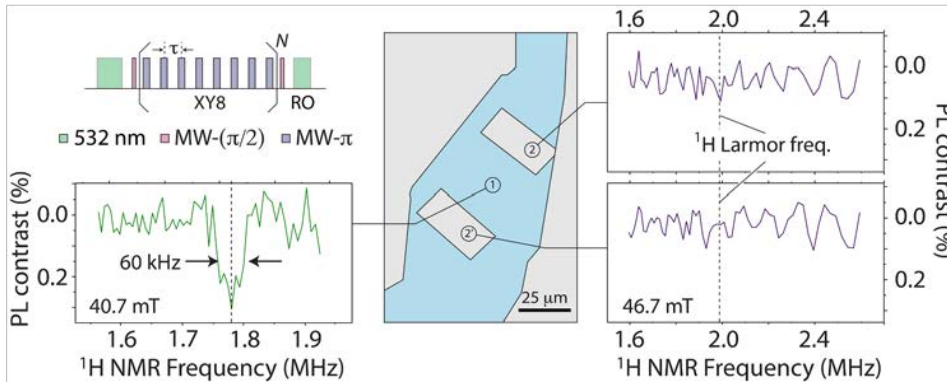

**Figure S4 | NV-NMR in and out of the channel area.** (Left) Pulse sequence and NV-NMR signal from a site in the channel area, Section ① (top and bottom, respectively). (Center) Schematic of the patterned hBN flake; ① and ② respectively denote the channel and window areas. (Right) We observe no $^1$H NV-NMR signal from either window in the structures (sections ②).

We divide our experiments into "measurement rounds", each starting with nanochannel exposure to water followed by extended observation periods, whose duration (typically two to three weeks) depends on the rate of water evaporation (noticeable despite chip encapsulation of the diamond crystal, Fig. S2). Figure 3a in the main text illustrates this point with the $^1$H NMR spectra observed immediately before and after a channel fill (marking the separation between two consecutive measurement cycles). To compensate for long-term system instabilities — created, e.g., by lab temperature fluctuations — we limit individual NMR data collection runs to 12–15 hour windows, which we intersperse with ODMR and Rabi measurements for regular MW frequency and amplitude monitoring; we also integrate an automated "site tracking" protocol to avoid laser beam drifting during prolonged experiments. For future reference, pulsed green optical excitation — with characteristic intensity of order 6 GW/m$^2$ in our diffraction-limited microscope — is invariably present throughout all these protocols for optimal SNR.

We limit our discussion to signals recorded by NVs within the channel area. Although water is expected to adsorb onto all diamond surfaces during the channel fill, we failed to observe $^1$H NMR signatures outside this section of the structure (Fig. S4). This could indicate the adsorbed water layer is just too thin under normal operating conditions and falls below our detection limit, though it could also be a consequence of the NMR resonance being too broad; this later case, in turn, does not necessarily imply high water diffusivity, as the broadening could originate from hyperfine interactions with surface charge (see below, Section II).

## II. Data analysis

### II.1 *Self-diffusivity of water under nano-channel confinement*

To characterize water diffusion, we implement NV correlation spectroscopy using two consecutive XY8-*N* pulse trains resonant with the $^1$H Larmor frequency. Reproduced in Fig. S5a, this protocol probes the auto-correlation function of the NV time evolution in the presence of the statistical magnetic field created by the sample nuclear spins. The stochastic nature of this nuclear "spin noise" makes the scheme sensitive to diffusion because any possible correlation between the phases picked up by a probe NV during each encoding segment vanishes if the source nuclear spins diffuse away from the sensing volume during the interval $\tilde{\tau}$ separating the pulse trains. The correlation signal takes therefore the form[5]

$$C(\tilde{\tau}) \propto \langle B_n^2 \rangle \cos(\omega_n \tilde{\tau}) F\left(\frac{\tilde{\tau}}{T_D}\right), \tag{1}$$

where $\langle B_n^2 \rangle$ is the root-mean-square field produced by the nuclear spins, $\omega_n$ is the nuclear Larmor frequency, and $F\left(\frac{\tilde{\tau}}{T_D}\right)$ is a decay envelope, where $T_D$ denotes the molecular self-diffusion time. The latter is connected to the sample self-diffusion coefficient $D$ via the relation $d \approx \sqrt{6DT_D}$, with $d$ representing the detection volume radius (in turn, approximately equal to the NV distance to the surface[7]).

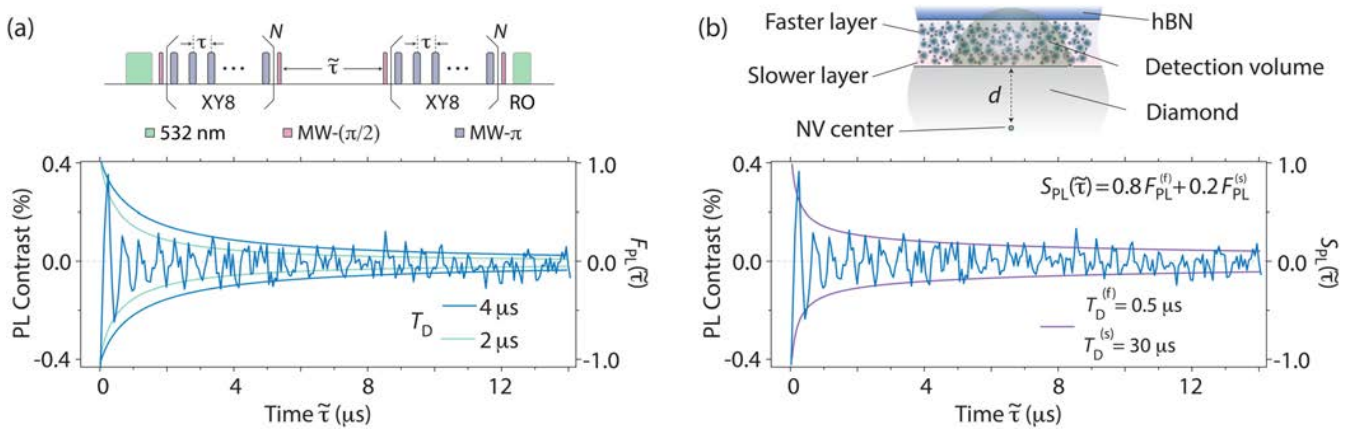

**Figure S5 | $^1$H correlation spectroscopy as a probe of water diffusion.** (a) Correlation protocol and $^1$H correlation signal (reproduced from Fig. 3b in the main text). The blue (green) envelope was calculated from Eq. (2) assuming $T_D = 4$ μs ($T_D = 2$ μs). Agreement is poor at short times when $T_D \gtrsim 4$ μs; we observe the converse if $T_D \lesssim 2$ μs. (b) (Top) We assume a two-layer model where water molecules proximal to the diamond surface move more slowly. (Bottom) Same as in (a) but assuming an envelope of the form $S_{PL}(\tilde{\tau}) = 0.8 F_{PL}^{(f)}(\tilde{\tau}) + 0.2 F_{PL}^{(s)}(\tilde{\tau})$ where $F_{PL}^{(f)}$ ($F_{PL}^{(s)}$) has characteristic time $T_D^{(f)}$ ($T_D^{(s)}$). We attain reasonable agreement both after short and long evolution intervals.

The decay envelope $F\left(\frac{\tilde{\tau}}{T_D}\right)$ has been often assumed to follow an exponential dependence with characteristic decay time $T_D$[6]. A more detailed analysis that takes into consideration the sensing geometry[8] — restricted by the host diamond to half a sphere — shows instead the response at long times $\tilde{\tau} \gg T_D$ has the form $F_{PL}\left(\frac{\tilde{\tau}}{T_D}\right) \propto \left(\frac{\tilde{\tau}}{T_D}\right)^{-\frac{3}{2}}$, where the label in $F$ stands for the predicted power law dependence. Compared to an exponential function, the slower power law decay better captures the long-lived shape of the nuclear spin correlation signals derived from observations of $^1$H spins in oil drops on the diamond surface[6,8].

The bottom half in Fig. S5a compares the present results with the predicted correlation envelope $F_{PL}$, whose complete functional dependence[8] we reproduce below for completeness, namely

$$F_{PL}(z) = \frac{4}{\sqrt{\pi}}\left(z^{-\frac{3}{2}} - \frac{3}{2}z^{-\frac{1}{2}} + \frac{\sqrt{\pi}}{4} + 3z^{\frac{1}{2}} - \frac{3\sqrt{\pi}}{2}z + \sqrt{\frac{\pi}{z}}\,\text{erfc}\left(z^{-\frac{1}{2}}\right)\exp(z^{-1})\left(-z^{-\frac{3}{2}} + z^{-\frac{1}{2}} - \frac{7}{4}z^{\frac{1}{2}} + \frac{3}{2}z^{\frac{3}{2}}\right)\right) \quad (2)$$

where $z \equiv \tilde{\tau}/T_D$. Indeed, the measured correlation signal does show a non-exponential envelope characterized by a quick decay at early times followed by a longer-lived coherence, but we find only qualitative agreement with Eq. (2) suggesting the diffusion dynamics of entrapped water is more complex. Super-imposed solid traces corresponding to Eq. (2) for different diffusion times — specifically, 2 and 4 μs — show longer values for $T_D$ better describe the response at long times but fail to capture the rapid decay seen for short evolution intervals. This behavior could stem from heterogeneity in the self-diffusion coefficient, longer for water layers closer to the surface. For example, we attain improved agreement with a simplified, two-layer model where $S_{PL}(\tilde{\tau}) = \alpha F_{PL}^{(f)}(\tilde{\tau}) + \beta F_{PL}^{(s)}(\tilde{\tau})$ with characteristic diffusion times $T_D^{(f)} = 0.5$ μs and $T_D^{(s)} = 30$ μs and normalized weights $\alpha = 0.8, \beta = 0.2$ (Fig. S5b). Using $d = 8$ nm as the characteristic size of the detection volume, we derive self-diffusion constants $D^{(f)} = d^2/6T_D^{(f)} \sim 21 \times 10^{-12}$ m$^2$ s$^{-1}$ and $D^{(s)} \sim 0.4 \times 10^{-12}$ m$^2$ s$^{-1}$, much below the self-diffusion coefficient of bulk water ($2.3 \times 10^{-9}$ m$^2$ s$^{-1}$). A similar analysis of the correlation signal in Fig. 4b of the main text — not shown here for brevity — indicates the relaxation time in this slower layer can exceed 100 μs, hence leading to $D \leq 0.1 \times 10^{-12}$ m$^2$ s$^{-1}$. Our analysis is qualitatively consistent with the heterogeneity observed for entrapped water by MD modeling (see below, Section III) where molecular diffusion is found to depend sensitively on the distance to the walls when charges are present (see also Figs. 4c through 4e in the main text).

**II.2 *Hyperfine interactions***

Inspection of the correlation signal's Fourier transform shows an asymmetric spectrum with maximum at the anticipated $^1$H Larmor frequency — 1.91 MHz at the applied magnetic field — but surrounded by shoulders spanning a ~2 MHz range (Fig. S6a). Since this broadening largely exceeds that possible via $^1$H chemical shifts or internuclear dipolar couplings, we theorize the observed pattern stems from a different spin interaction, possibly hyperfine in nature.

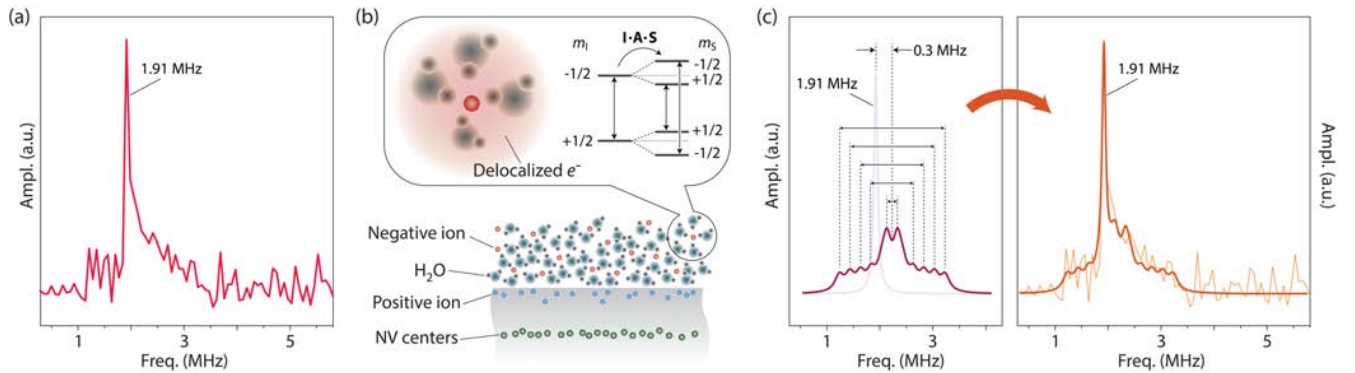

**Figure S6 | Hyperfine-induced broadening?** (a) Fourier transform of the $^1$H spin correlation signal as shown in Fig. S5. Besides the main nuclear spin resonance at the $^1$H Larmor frequency, the spectrum displays an asymmetric broadening spanning ~2 MHz. (b) Photogenerated carriers lead to the formation of an electric double layer, with positive charge likely accumulating at the interface and electrons solvating into the fluid. We hypothesize hyperfine couplings between near-surface $^1$H nuclei and the electronic spins of these carriers — respectively denoted by vector operators **I** and **S** — lead to splitting of the proton NMR resonance with amplitude ||**A**||. Arrows on the energy diagram (right hand side insert) indicate allowed transition frequencies. (c) We describe the $^1$H NMR spectrum as a superposition of contributions from water structures experiencing different hyperfine couplings (here discretized to take five different values). The combination of hyperfine-shifted and hyperfine-free contributions lead to a pattern consistent with experiment. Note the positive ~0.3 MHz shift of the hyperfine pattern center frequency, suggestive of strong non-diagonal contributions to the hyperfine tensor.

Supporting this hypothesis is the recent observation of charge injection from shallow NVs and coexisting surface traps under green excitation[9,10], here activated by the combination of day-long measurement times and the insulating nature of all components in the nanochannel structure (deprived from electrical grounding and hence prone to the formation of space charge fields). The injection of solvated electrons, not holes, seems the most likely scenario given the negative electron affinity of diamond surfaces in contact with water[11] (Fig. S6b).

To model the interaction between $^1$H spins and the excess electron in a given molecular cluster, we write the Hamiltonian as

$$H = -\omega_n I_z + A I_z S_z + A' I_x S_z. \tag{1}$$

In the above expression, $\omega_n$ is the proton Larmor frequency at the applied magnetic field, **I** and **S** are vector operators respectively representing the $^1$H and electron spins, $A$ and $A'$ denote hyperfine coupling constants, and we assume $\hbar = 1$. In the representation corresponding to the eigenstates of $I_z$ and $S_z$, the Hamiltonian takes a block-diagonal form; for the $|m_S = -1/2\rangle$ manifold, the eigenenergies are given by

$$E_\pm^{(-1/2)} = \pm\sqrt{\frac{1}{4}\left(\omega_n + \frac{A}{2}\right)^2 + \frac{A'^2}{16}} \approx \pm\frac{1}{2}\left(\omega_n + \frac{A}{2} + \frac{A'^2}{8\omega_n}\right), \tag{2}$$

where we assume $A \cong A' < \omega_n$ (see energy diagram in Fig. S6b). Similarly, we find for the $|m_S = +1/2\rangle$ manifold

$$E_\pm^{(+1/2)} = \pm\sqrt{\frac{1}{4}\left(\omega_n - \frac{A}{2}\right)^2 + \frac{A'^2}{16}} \approx \pm\frac{1}{2}\left(\omega_n - \frac{A}{2} + \frac{A'^2}{8\omega_n}\right). \tag{3}$$

Since the NMR-allowed transitions must preserve the electron spin projection, we find the resonance frequencies in the hyperfine doublet take the values

$$\omega_n^\pm = \omega_n \pm \frac{A}{2} + \frac{A'^2}{8\omega_n}. \tag{4}$$

Equation (4) implies that in the presence of hyperfine couplings, the $^1$H resonance splits into two peaks, each separated from the center frequency by $A/2$. Note that the average frequency $(\omega_n^+ + \omega_n^-)/2$ shifts from $\omega_n$ by a positive amount $A'^2/8\omega_n$; as an example, for the present magnetic field conditions ($\omega_n \sim 2$ MHz) and assuming $A' \approx 1$ MHz, we obtain $A'^2/8\omega_n \approx 60$ kHz.

Now, we can tentatively rationalize the $^1$H spectrum as the result of heterogeneous contributions from protons experiencing hyperfine couplings varying over a ~2 MHz range, each group manifesting as a spectral doublet (whose center is potentially blue-shifted from the $^1$H Larmor frequency). To prove this point, we choose five different hyperfine couplings with amplitudes $A_i$, $i = 1 \cdots 5$; in all cases we use $A'_i \sim 2$ MHz, impose a linewidth of 0.1 MHz, and select varying amplitudes to create the hyperfine pattern shown on the left panel of Fig. S6c. Assuming the inner water layers experience negligible hyperfine splitting, we can qualitatively reproduce the observed spectrum if we supplement the hyperfine shifted pattern with a resonance peak at the $^1$H Larmor frequency (right-hand side panel in Fig. S6c). Note that the same rationale can be extended to reproduce the bottom spectrum in the series of Fig. 4c of the main text, except that in this case, the observation of resolved satellites suggests water molecules and carriers organize to adopt preferred structures featuring specific hyperfine coupling constants.

Notably, the hyperfine strengths we derive have order of magnitude comparable to that extracted for protons in the solvation layer of ions dissolved in bulk water as measured via inductive liquid-state NMR[12]. We warn, however, the above assignments should only be seen as a crude guide since the varying relative amplitudes of the resonance pairs and limited signal-to-noise ratio (SNR) combine to create some ambiguity. More importantly, while H$_2$O molecules are known to coordinate around solvated electrons, the resulting clusters diffuse rapidly[13] and are thus difficult to reconcile with our observations without also imposing temporal stability over a time scale comparable to the duration of our measurements. Lastly, we emphasize different measurement runs — understood as the set of experiments during the time interval separating two consecutive channels fills, typically 2 to 3 weeks — do not necessarily yield the exact same evolution in the $^1$H NMR spectra we presented in Fig. 4c of the main text, perhaps a consequence of the varying nature of the space charge fields forming under laser excitation. This variability could also originate from heterogeneities in the channel structure and/or surface termination, as residual NV PL changes between different rounds complicate our ability to replicate the same (diffraction-limited) location for measurements belonging to different runs.

## III. Molecular dynamics simulations

We perform classical molecular dynamics (MD) simulations of a system composed of $N \sim 30000$ water molecules, with and without additional charges, sandwiched by an hBN monolayer and a diamond crystal as described in the main text. We consider four diamond surfaces terminated with either (i) H, (ii) O, (iii) OH, and (iv) a combination of H and OH groups (arranged periodically over the surface). Water molecules are represented using the TIP4P/2005 water model[14]. During the MD simulations, the hBN and diamond atoms are immobile except for the H atoms belonging to the diamond OH groups (cases (iii) and (iv)). The distance between the hBN monolayer and the outmost C atoms of diamond is set to ~5.771-5.856 nm depending on the diamond surface considered. The hBN/diamond surfaces are perpendicular to the $z$-axis and have an area $A = \Delta y \cdot \Delta x$ with $\Delta y = 13.137 - 13.197$ nm and $\Delta x = 13.137 - 13.197$ nm depending on the surface considered (see below). The diamond crystal thickness is $\Delta z = 1.151 - 1.226$ nm (depending on the diamond surface termination). Increasing the values of $\Delta z$ does not affect our results because the additional C atoms are located at least 1.15 nm apart from any water molecule in the confined volume and the corresponding C-water interactions are null (the Lennard-Jones interactions between the water molecules and the diamond C atoms are short-ranged, and are evaluated within a cutoff distance $r_c = 1.10$ nm). All MD simulations are performed at constant temperature ($T = 300$ K) and volume. The system box side lengths are ($L_z = 30.000, L_y = \Delta y, L_x = \Delta x$) and periodic boundary conditions apply to all three directions (note that $L_z$ is much larger than the separation between the plates (<6 nm) and the diamond thickness $\Delta z$; this extra empty space is left to minimize any potential effect that may be due to the periodic boundary conditions along the $z$-direction).

All computer simulations are performed using the GROMACS software package (2022.6-spack)[15]. The system temperature is maintained using a Nose-Hoover thermostat with a time constant of 0.5 ps[16,17]. Coulombic interactions are evaluated using the particle mesh Ewald technique[18] with a Fourier spacing of 0.12 nm. For both the Lennard-Jones (LJ) and the real space Coulomb interactions, we use a cutoff of $r_c = 1.10$ nm. To control the water OH covalent bonds lengths, we use the LINCS (Linear Constraint Solver) algorithm. MD simulations are performed for 8 ns which is long compared with the relaxation time of bulk TIP4P/2005 water ($\tau \sim 50–100$ ps) at $T = 300$ K and $P = 0.1$ MPa[19] (under these conditions, it takes <100 ps for the mean-square displacement of the water molecules to be $> 1$ nm$^2$). The simulation time step is $dt = 1$ fs.

### III.1 *Systems preparation*

For each diamond surface studied, we carry out three kinds of MD simulations as follows:

*A)* We first perform MD simulations of a thick film of water (~6.5 nm-thick; $N = 33628$) covering the whole diamond surface (Fig. S7a). The system is simulated for 4 ns allowing for water to reach equilibrium. After a few ps, water molecules at the diamond interface rearrange into three layers (red line in Fig. S7c). These water layers are separated from the liquid-vapor interface (at $z > 5.5$ nm) by a ~4 nm-thick volume of bulk-like water. We note that this bulk-like water is at $P \sim 0$ MPa since it is in contact with a vapor volume and the average liquid vapor interface is flat. Accordingly, the density of the thick bulk-like water volume is independent of the surface considered; in all cases we find that the density of the bulk-like water is close to $\rho_{bulk} \sim 0.997$ g/cm$^3$ which is the density of bulk TIP4P/2005 water at $P = 0.1$ MPa and $T = 300$ K[14] (see Fig. S7c).

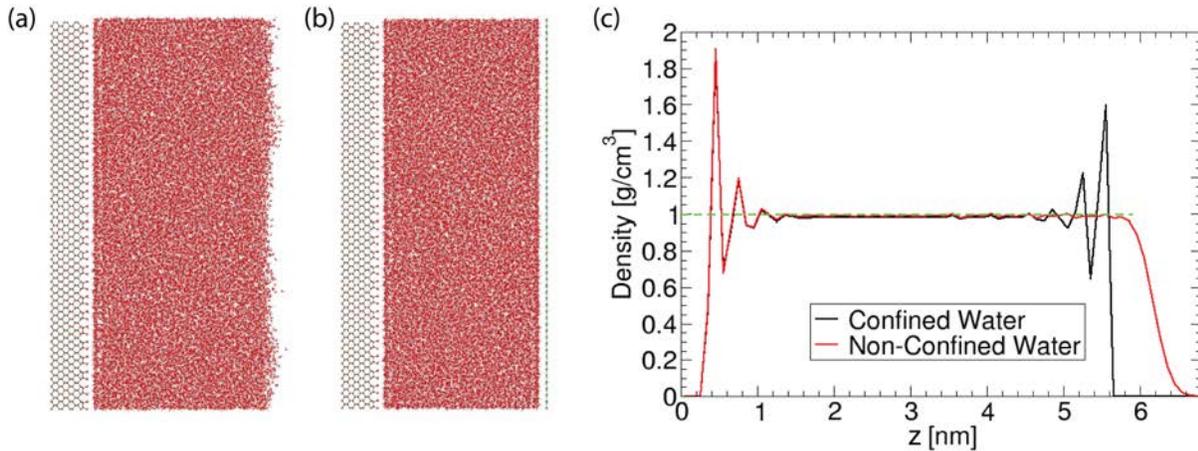

**Fig. S7 | System preparation.** (a) Snapshot from MD simulations of a ~6.5 nm-thick water film on a diamond crystal passivated with OH groups. The water film is separated from the vapor phase by a flat interface that fluctuates during the MD simulation. (b) Snapshot from MD simulations of the system shown in (a) after adding a hBN monolayer to confine the water volume. (c) Density profile of water as a function of the distance $z$ from the diamond surface for the systems shown in (a) and (b). In both cases, three water layers form near the diamond interface, and a bulk-like volume of water exists at approximately $z > 1.5$ nm. In system (a), red line, the density profile decays smoothly to zero as the vapor phase is reached; in system (b), three water layers from next to hBN. The diamond-hBN separation in system (b) is tuned to match the density of bulk TIP4P/2005 water at $z \sim 2 - 4$ nm (the horizontal dashed line in (c) indicates the density of bulk TIP4P/2005 water at $P = 0.1$ MPa and $T = 300$ K, $\rho_{\text{bulk}} \sim 0.997$ g/cm$^3$).

B) A second family of MD simulations is then carried out by adding an hBN monolayer at a distance $d = 5.771 - 5.856$ from the diamond surface (depending on the diamond surface termination, Fig. S7b). The distance $d$ is selected so the density of the confined water in the intermediate region between the hBN/diamond surfaces matches the density of bulk (TIP4P/2005) water, $\rho_{bulk} \sim 0.997$ g/cm$^3$ (see Fig. S7c).

C) In order to include negative charges in the confined water volume, we proceed as follows. For a target diamond surface, we take an equilibrated configuration of the corresponding diamond-water MD simulation described in (B). Then, we replace randomly $N_q$ water molecules by fluoride anions. The fluoride anions are modeled using the Madrid2019 force field[20] where each anion has an effective charge $q = -0.85\ e$. Hence, to make the system charge-neutral, we also add a total positive charge $N_q|q|$ to the diamond crystal. This charge is distributed evenly over the C monolayer of the diamond crystal that is located closest to the confined water (these C atoms form a covalent bond with the surface O and are shown in blue in Fig. 4d of the main text). This C monolayer contains $N_c$ atoms ($N_c = N_q = 2704$ in the case of diamond crystals passivated with H, OH, and both H/OH groups; $N_c = 2738$ for the diamond terminated with O groups). We note that replacing the water molecules by fluoride leads to a redistribution of water molecules and to a decrease in the overall density of water in the confined volume. We will later see this can lead to the formation of vapor-like domains (see below Fig. S13a). We note that reducing the separation between the hBN monolayer and diamond suppresses the formation of vapor domains while leaving our results unaffected.

### III.2 *Model surfaces for Diamond and hBN*

*Diamond.* Fig. S8 shows side views of the four diamond surfaces considered. The C100 surface of diamond is exposed to the water molecules, and it is passivated with either (a) O, (b) H, (c) OH, or (d) H and OH groups. The diamond atoms are maintained at fixed positions except for the H atoms of the hydroxyl groups (cases (c) and (d)). These H atoms are mobile, but they are constrained by (i) an OH stretching bond potential of the form

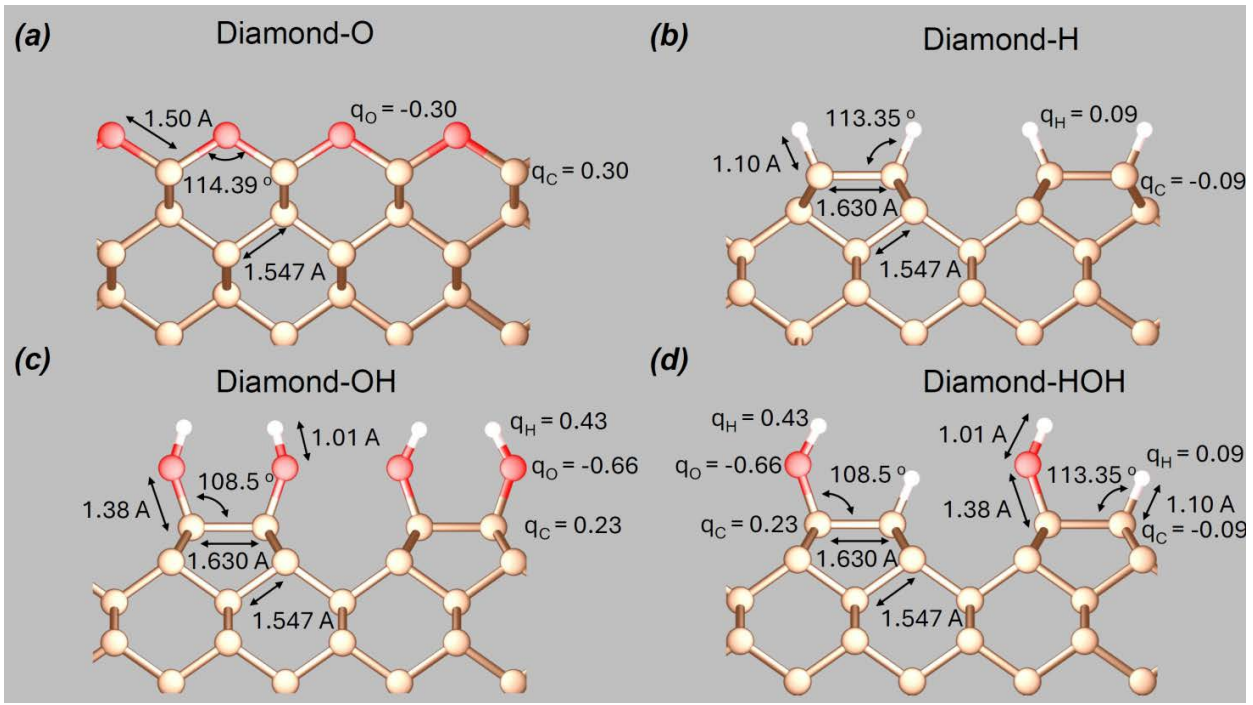

**Fig. S8 | Structure of the diamond surfaces**. Termination with (a) O, (b) H, (c) OH, and (d) both H and OH groups. The bond lengths, angles, and partial charges are indicated; LJ interaction parameters are given in Tables 1-4. The partial charges of the O and H atoms are adapted from Ref. [21]; the partial charges of the C atoms forming a covalent bond with the surface O and H atoms are such that the total charge of the C-H and C-O-H groups are zero.

$$V_{\text{OH}}(r) = \tfrac{1}{2} k_r (r - r_0)^2, \tag{5}$$

where $r$ is the OH distance, and (ii) by an HOC angle potential given by

$$V_{\text{HOC}}(\theta) = \tfrac{1}{2} k_\theta (\theta - \theta_0)^2. \tag{6}$$

Parameters $k_r$, $k_\theta$, $r_0$, and $\theta_0$ are adapted from the CHARMM force field for alcohols and carbohydrates[21], specifically, $k_r = 456056.00$ kJ/mol/nm, $k_\theta = 481.160$ kJ/mol/rad$^2$, $r_0 = 0.09600$ nm and $\theta_0 = 106.00°$. The structure of the diamond crystals terminated with H, O, and OH groups (Figs. S8a-c) are adapted from Refs. [22, 23]; the structure of the diamond crystal passivated with both H and OH groups (Fig. S8d) is a model surface that we created based on the diamond surfaces terminated with H and OH groups (Figs. S8b and S8c). The covalent bond lengths and bond angles among the immobile atoms are indicated in the figures. Only the H and O atoms at the surface, and the C atoms forming a covalent bond with the O and H atoms have a non-zero partial charge; the corresponding partial charges are also indicated in the figures. The LJ parameters $\varepsilon$ and $\sigma$ for the atoms in the different walls are given in Tables S1-S4. The Lorentz-Berthelot combination rules are used to generate the LJ parameters among the diamond atoms and the anions/water atoms.

Table S1. LJ Params for the diamond crystal terminated with O atoms.

| Atom type | $\sigma$ [nm] | $\varepsilon$ [kJ/mol] |
|---|---|---|
| O (in C-O) [a] | 0.293997 | 0.50208 |
| C (in C-O) [b] | 0.321400 | 0.23633 |
| C [b] | 0.321400 | 0.23633 |

[a] Adapted from Ref. [21]; [b] From Ref. [24].

Table S2. LJ Params for the diamond crystal terminated with H atoms.

| Atom type | $\sigma$ [nm] | $\varepsilon$ [kJ/mol] |
|---|---|---|
| H (in C-H) [a] | 0.235200 | 0.09205 |
| C (in C-H) [b] | 0.321400 | 0.23633 |
| C [b] | 0.321400 | 0.23633 |

[a] Adapted from Ref. [21]; [b] From Ref. [24].

Table S3. LJ Params for the diamond crystal terminated with OH groups.

| Atom type | $\sigma$ [nm] | $\varepsilon$ [kJ/mol] |
|---|---|---|
| H (in C-O-H) [a] | 0.040001 | 0.19246 |
| O (in C-O-H) [a] | 0.302906 | 0.63639 |
| C (in C-O-H) [b] | 0.321400 | 0.23633 |
| C [b] | 0.321400 | 0.23633 |

[a] Adapted from Ref. [21]; [b] From Ref. [24].

Table S4. LJ Params for the diamond crystal terminated with H and OH groups.

| Atom type | $\sigma$ [nm] | $\varepsilon$ [kJ/mol] |
|---|---|---|
| H (in C-O-H) [a] | 0.040001 | 0.19246 |
| O (in C-O-H) [a] | 0.302906 | 0.63639 |
| C (in C-O-H) [b] | 0.321400 | 0.23633 |
| H (in C-H) [a] | 0.235200 | 0.09205 |
| C (in C-H) [b] | 0.321400 | 0.23633 |
| C [b] | 0.321400 | 0.23633 |

[a] Adapted from Ref. [21]; [b] From Ref. [24].

*Boron nitride.* The structure of hBN is hexagonal with the B and N atoms kept in fixed positions throughout the entire MD simulations. The experimental BN covalent bond distance is ~1.46 A. However, in order to make the system periodic along the $y$-and $x$-directions, we stretched slightly the hBN monolayer depending on the diamond surface considered (which is periodic along the $y$-and $x$-directions). Figure S9a shows the geometry of the hBN monolayer in the presence of the diamond crystal terminated with O atoms. The geometry of hBN in the presence of the diamond crystal terminated with H, OH, and H+OH groups is shown in Fig. S9b. The N/B atoms interact with the water molecules and anions via electrostatic and LJ interactions; partial charges and LJ interaction parameters are given in Table S5. The Lorentz-Berthelot combination rules are used to generate the LJ parameters among the hBN atoms and the anions/water atoms.

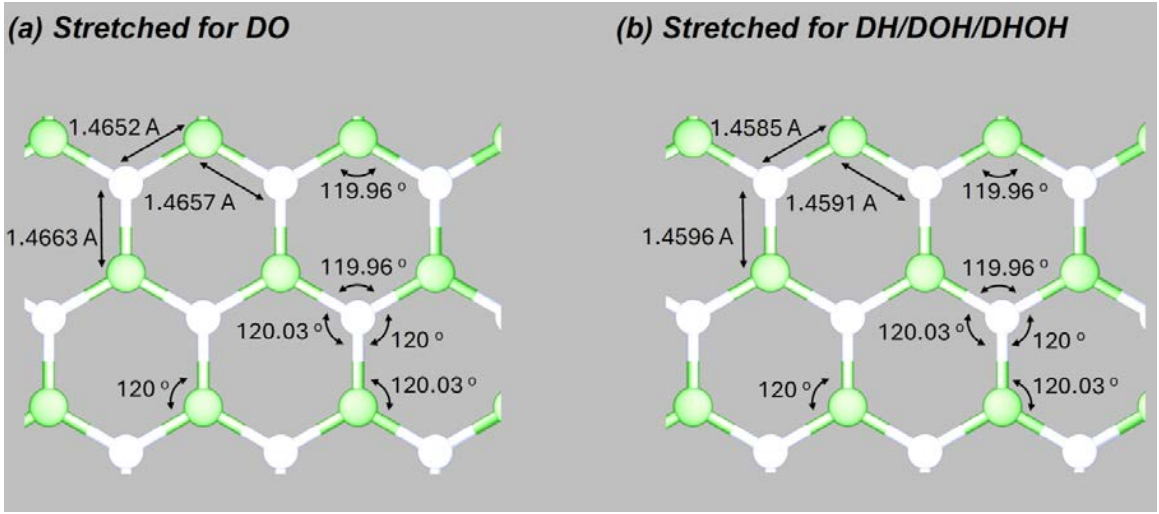

**Fig. S9 | Structure of the hBN monolayer.** The bond lengths and angles are indicated. Partial charges and LJ interaction parameters are given in Table S5.

Table S5. LJ parameters and partial charges for hBN.

| Atom type | σ [nm] | ε [kJ/mol] | q [e] |
|---|---|---|---|
| B [a] | 0.330870 | 0.28970 | 0.907 |
| N [a] | 0.321740 | 0.19790 | -0.907 |

[a] From Ref. [25].

### III.3 *Impact of the charge distribution on the diamond surface*

The results reported in the main text are based on a confined water system containing $N_q = 2704$ anions (fluoride). As explained below, to compensate for the negative charges added to the confined water, we also *add* a positive charge to the C atoms located in the monolayer closest to the water volume. These C atoms form a covalent bond with the surface O atoms and are shown in blue in Fig. 4d of the main text; see also Fig. S10a below. In the case of the diamond surface terminated with OH groups, there are $N_c = N_q$ atoms in the C monolayer closest to water. Hence, each of these C atoms (shown in blue in Fig. S10a) are assigned an additional charge $q' = -q > 0$. Below, we study how the results from our MD simulations vary when the net positive charge in diamond is distributed uniformly over *all* the C atoms of the diamond crystal.

Figure S10a shows a snapshot of the same system studied in the main text (Fig. 4c-e; briefly, System 1) but with the compensating positive charge of diamond distributed over *all* C atoms of the diamond crystal, as explained above (briefly, System 2). Figure S10b shows the density profiles of the F⁻ ions and water taken from the main text (Fig. 4c; System 1) and the corresponding density profiles when the compensating positive charge is distributed over *all* C atoms in the diamond (System 2). The density profiles of Systems 1 and 2 practically overlap with one another. Figure S10c shows the fraction of water molecules in layers L1, L2, and L3 as a function of time. Again, these distributions are practically identical for Systems 1 and 2. Therefore, our results are insensitive to the specific distribution of the added positive charge over the diamond crystal.

The results in Figs. S10b and S10c can be understood by approximating each C monolayer within the diamond crystal with a homogeneously charged, infinite plane. Since an infinite charged plane generates a constant electric field (perpendicular to the plane), the same net electric field is generated if the compensating $N_q|q|$ charge in diamond is placed in a single C monolayer or multiple C monolayers.

### III.4 *Impact of diamond surface termination*

The results presented in Figs. 4c-e of the main text are based on a system comprising $N = 27738$ water molecules and $N_q = 2704$ fluoride atoms confined by hBN and hydroxylated diamond. Similar results hold when diamond is terminated with (i) O, (ii) H, and (iii) both H and OH groups. To show this, we include in Figs. S11, S12, and S13 the density profiles for cases (i), (ii), and (iii), respectively. Also included is the fraction of water molecules that are initially in the first, second, and third layers next to the diamond crystal that remain in the corresponding layer at time $t$ (results for the hydroxylated diamond surface are in Fig. S10c). In all cases, the fluoride anions are located preferentially next to the diamond surface. The water molecules and fluoride ions next to the diamond surface are found at specific locations on the diamond structure, forming a square lattice. In this lattice, the fluoride ions and the first monolayer of water molecules are initially placed in a disordered manner, next to one another. Importantly, these water molecules and anions are unable to diffuse (cases (i) and (ii); see Figs. S11d and S12d) or have a very small diffusivity (case (iii); see Fig. S13d).

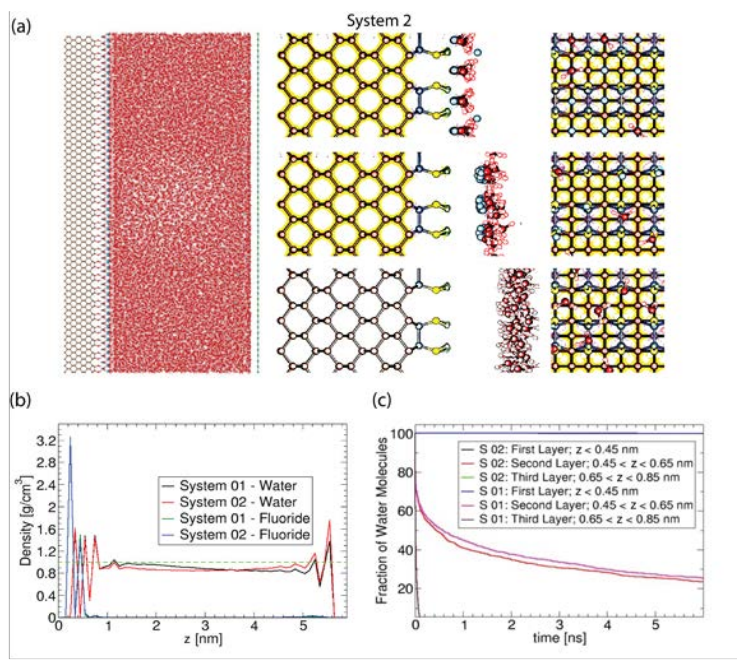

**Fig. S10 | Impact of the positive charge distribution in the diamond.** (a) Snapshot taken from our MD simulations of water ($N_w = 27738$) and fluoride anions ($N_q = 2704$) confined by hBN and diamond passivated with OH atoms. The hBN/diamond separation is $d \sim 5.7$ nm and $T = 300$ K (same as in Fig. 4a of the main text; briefly, System 1). To make the system charge-neutral, a net positive charge is distributed evenly over *all* C atoms of the diamond crystal (shown in blue and gray, denoted System 2). (b) Water and fluoride density profiles for Systems 1 and 2. (c) Fraction of water molecules that are initially in layers L1, L2, L3 and remain in the same layer at time $t$. The results in (b) and (c) are practically identical for Systems 1 and 2.

### I.5 *Nanoconfined water (no anions added)*

As discussed in the main text (Figs. 4a-b), the layer of non-diffusive water molecules next to the diamond surface is no longer present when the fluoride anions are removed. Under these conditions, water remains in the liquid phase with no signs of vapor domains (see Fig. 4a of the main text, reproduced in Fig. S14a below). While water molecules still form layers next to the diamond surface and hBN (see Fig. 4b of the main text, reproduced in Fig. S14c below), these molecules can diffuse into the confined volume, and *vice versa*. This is indicated by (i) the mean-square displacement of the water molecules included in Fig. 4b of the main text, and (ii) by the fraction of water molecules in layers 1, 2, 3 at time $t = 0$ (next to the diamond surface) that remain in the same layer at time $t$ (Fig. S14d),

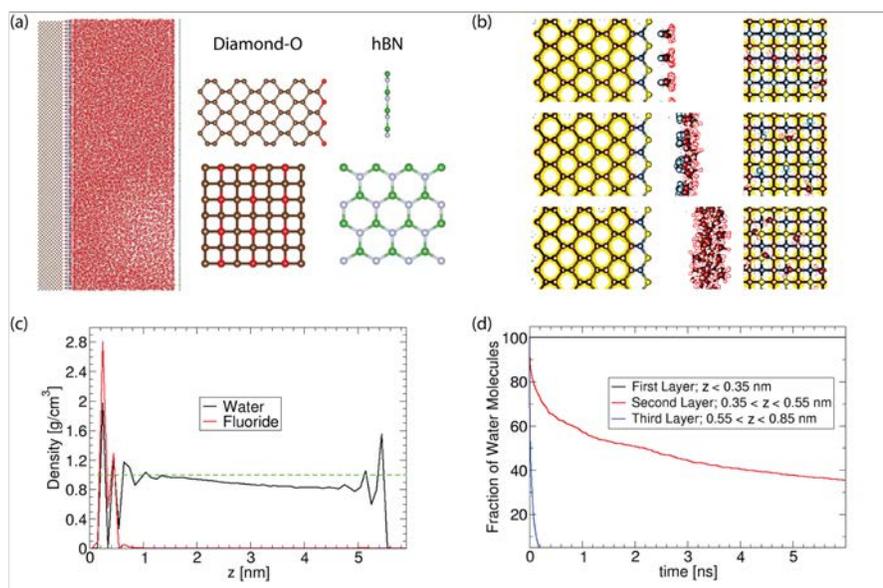

**Fig. S11 | O termination.** Snapshot from MD simulations of a system comprising $N_w = 27738$ water molecules and $N_q = 2704$ fluoride anions confined by an hBN monolayer and a diamond crystal passivated with O atoms. (b) Side and top view of the fluoride and water molecules located in the first (L1; top), second (L2, middle) and third (L3, bottom) water layers formed next to the diamond surface. (c) Water density profile as a function of the distance $z$ from the diamond surface showing three water layers (L1, L2, L3) next to the diamond surface. (d) Fraction of water molecules that are initially in layers L1, L2, and L3 that remain in the corresponding layer at time $t$. Water layers located in layer L1 are unable to diffuse.

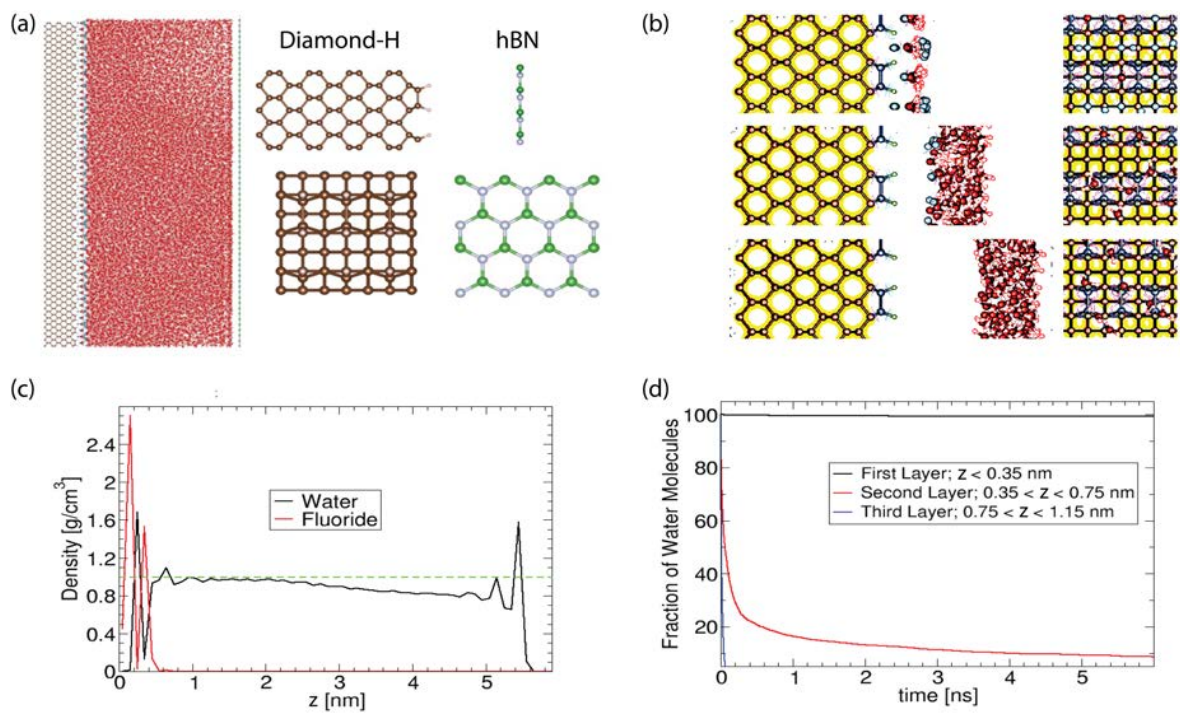

**Figs. S12 | H termination.** Same as in Fig. S11 for the case where the diamond surface is passivated with H atoms.

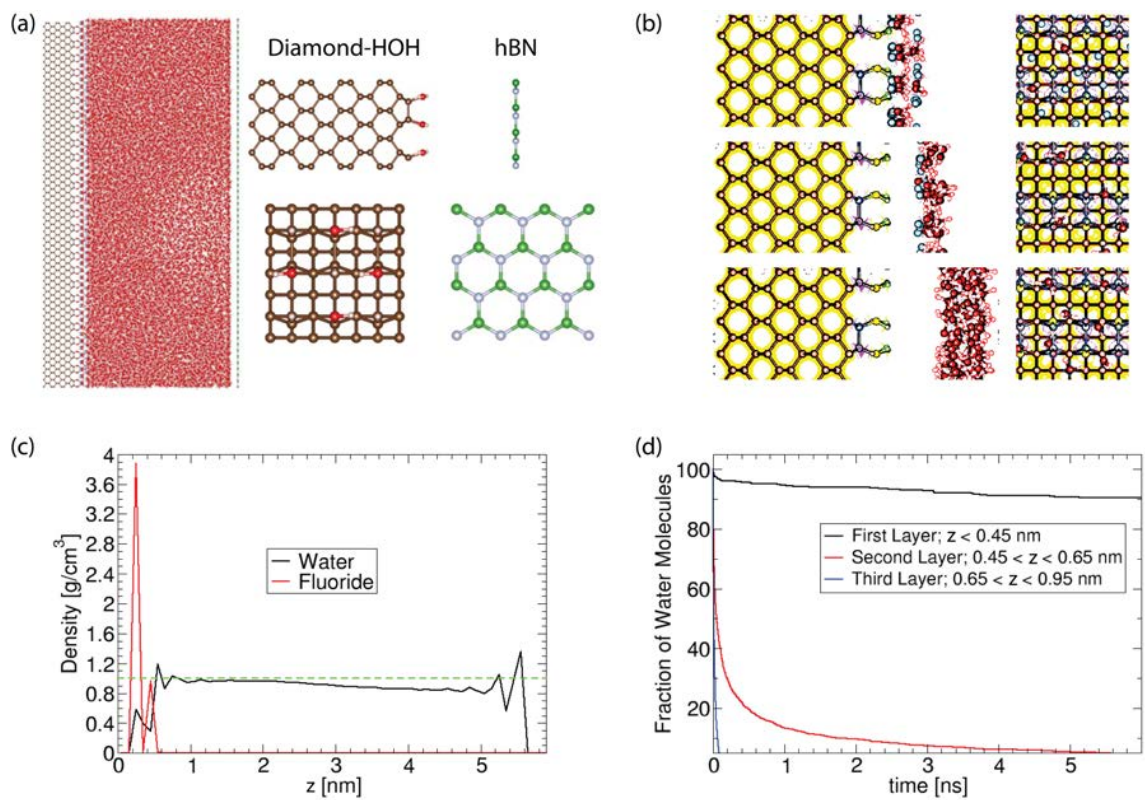

**Figs. S13 | H and OH termination.** Same as in Fig. S11 for the case where the diamond surface is passivated with H and OH groups.

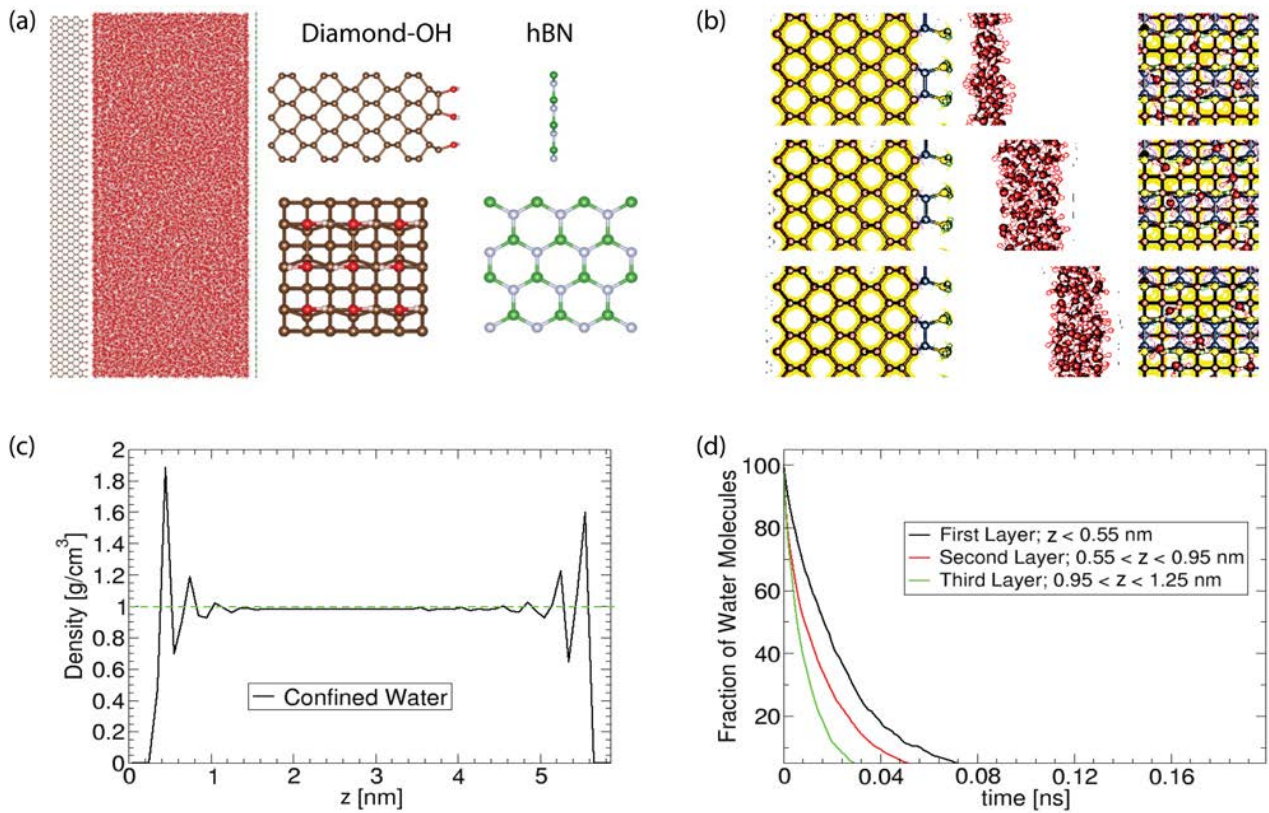

**Fig. S14 | Dynamics of pure water.** (a) Snapshot from MD simulations of water confined by an hBN monolayer, and diamond terminated with OH groups ($N = 30442$). The diamond crystal and hBN monolayer, as well as the diamond-hBN separation are identical to the system studied in Fig. 4 of the main text; $T = 300$ K. (b) Side and top view of the water molecules located in the first (L1; top), second (L2, middle) and third (L3, bottom) water layers formed next to the diamond surface. (c) Water density profile as a function of the distance $z$ from the diamond surface showing the three water layers (L1, L2, L3) formed next to the diamond surface. At approximately >1 nm away from the diamond/hBN, water exhibits bulk-like properties; the density of water in this region is practically identical to the density of bulk (TIP4P/2005) water at $P = 0.1$ MP and $T = 300$ K (dashed green line). (d) Average fraction of water molecules in layers L1, L2, L3 at time $t = 0$ (next to the diamond surface) that remain in the same layer at time $t$. Layers L1, L2, L3 are defined from (c) and correspond to $z < 0.55$ nm, $0.55 < z < 0.95$ nm, and $0.95 < z < 1.25$ nm, respectively.

### III.6 *Impact of the number of anions in the confined water*

To explore the effect of increasing the number of anions in the confined water, we perform MD simulations for the same system studied in the main text but with twice the number of fluorides in the system ($N_q = 5408$, Fig. S15a). As explained in the main text, the additional negative charge is compensated by adding a positive charge to the C monolayer of diamond closest to the water volume (blue C atoms in Fig. S15b). As shown in Fig. S15b-c, the anions are located preferentially next to the diamond surface and are surrounded by water molecules. The water molecules and anions next to the surface are located on a square lattice, as in the original system (compare Fig. 4d of the main text and Fig. S15b below); these water molecules are not able to diffuse (Fig. S15d). In this regard, our results are insensitive to the number of fluoride anions in the system ($N_{F^-} = 2704 - 5408$). The main effect of adding anions seems to be to alter the layering of water molecules next to the diamond surface. Specifically, as shown in Figs. S15b and S15c below, adding anions leads to an accumulation of negative charges at $0.40 < z < 0.6$ nm. Water molecules at these distances are mostly excluded.

### III.7 *Impact of anions size*

Our model system is based on classical MD simulations, and one may wonder if the results based on fluoride are sensitive to the anions size. From a classical perspective, an electron is represented by a point charge, much smaller and lighter than fluoride (while the effects due to the electron delocalization should play an important role, such effects cannot be included in our classical model system). Here, we compare the results reported in the main text when the fluoride mass is reduced by 94.7% (from $m_F = 18.998$ amu $m' = 1.008$ amu) and the Lennard-Jones (LJ) parameter σ, which defines the anion size, is reduced by 30 % (from $\sigma_F = 0.379$ nm to $\sigma' = 0.265$ nm); the anion charge and LJ parameter ε are left unchanged. Our results based on the modified fluoride anions are shown in Fig. S16. At the diamond-hBN separation studied, and for the given number of water and anions, a pronounced cavitation develops within the system (Fig. S16a). This is because of two effects: (i) The smaller modified anions move deeper within the groves of the diamond surface which (ii) allows water molecules to locate closer to the diamond surface (Fig. S16b-c). Both (i) and (ii) create additional space for the liquid to fill, hence, forcing the liquid to cavitate. Interestingly, despite the cavitation (next to the hBN monolayer), the molecules in the first water monolayer seem to be unable to diffuse (Fig. S16d), as found in the main text for the case of water with fluoride anions.

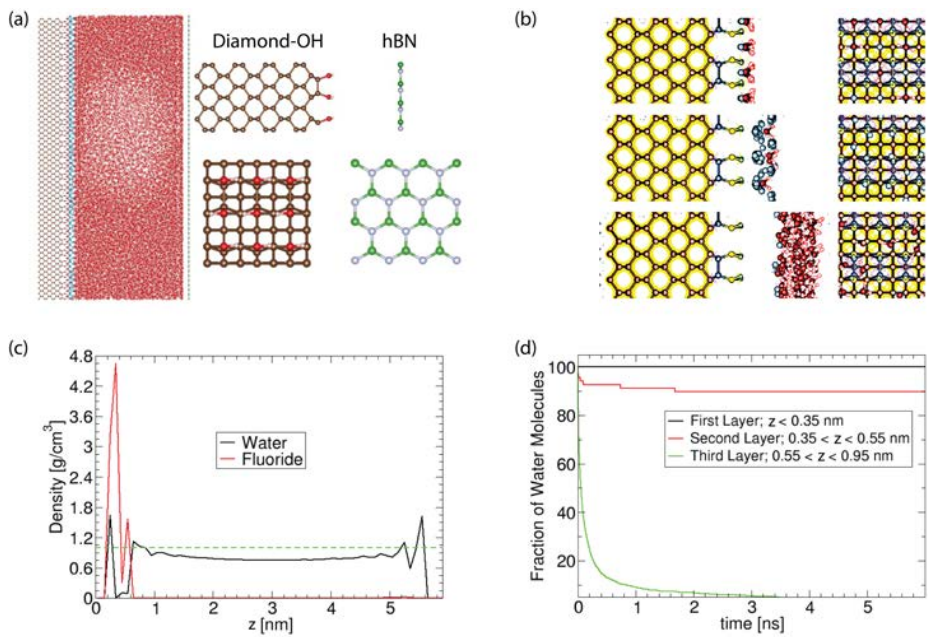

**Fig. S15 | Impact of the number of anions.** (a) Snapshot from MD simulations of a system comprising $N_w = 25034$ water molecules and $N_q = 5408$ fluoride anions confined by an hBN monolayer and a hydroxylated diamond crystal. (b) Side and top view of the fluoride and water molecules located in regions L1 (first water layer L1, $z < 0.35$ nm; top panel), L2 ($0.35 < z < 0.55$ nm; middle panel) and L3 (second water layer, $0.55 < z < 0.95$ nm; bottom panel). (c) Water density profile as a function of the distance $z$ from the diamond surface showing the three regions L1, L2, and L3 close to the diamond surface. (d) Fraction of water molecules that are initially in regions L1, L2, and L3 that remain in the corresponding region at time $t$. Water layers located in layer L1 are unable to diffuse.

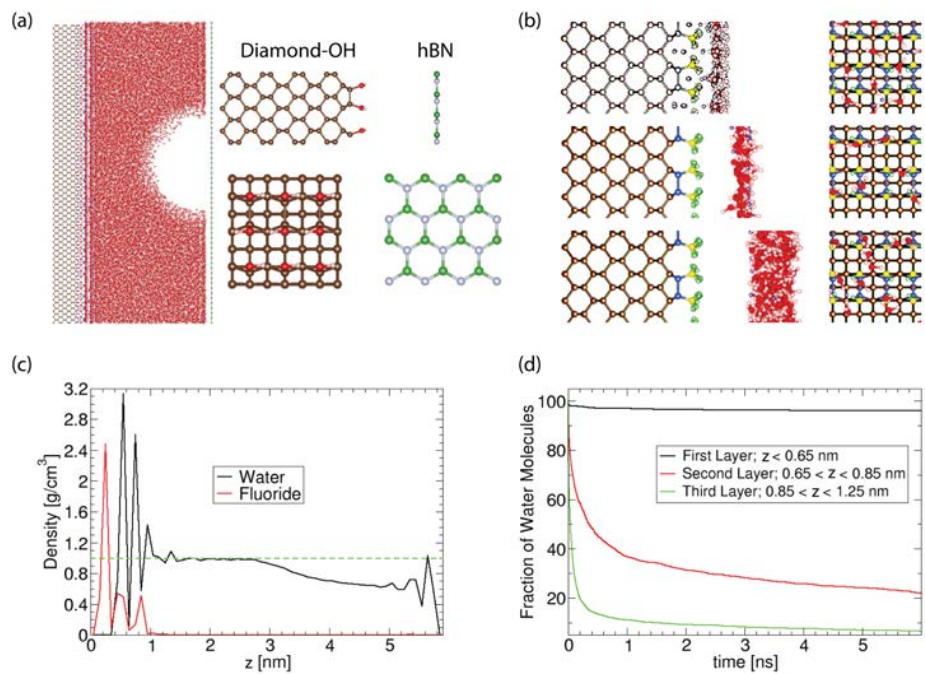

**Fig. S16 | Impact of the anion size.** Same system considered in Fig. 4c-e of the main text when the mass of the F- anions is reduced by 94.69% (from $m_F = 18.998$ amu $m' = 1.008$ amu) and the Lennard-Jones parameter σ is reduced by 30 % (from $\sigma_F = 0.379$ nm to $\sigma' = 0.265$ nm). (a) Snapshot of the system from the MD simulations. (b) Side and top view of the fluoride and water molecules located in the first ($z < 0.65$ nm, L1; top panel), second ($0.65 < z < 0.85$ nm, L2; middle panel), and third water layer ($0.85 < z < 1.25$ nm, L3; bottom panel). (c) Density profile as a function of the distance $z$ from the diamond surface showing the three regions L1, L2, and L3 close to the diamond surface. (d) Fraction of water molecules that are initially in water layers L1, L2, and L3 that remain in the corresponding region at time $t$.

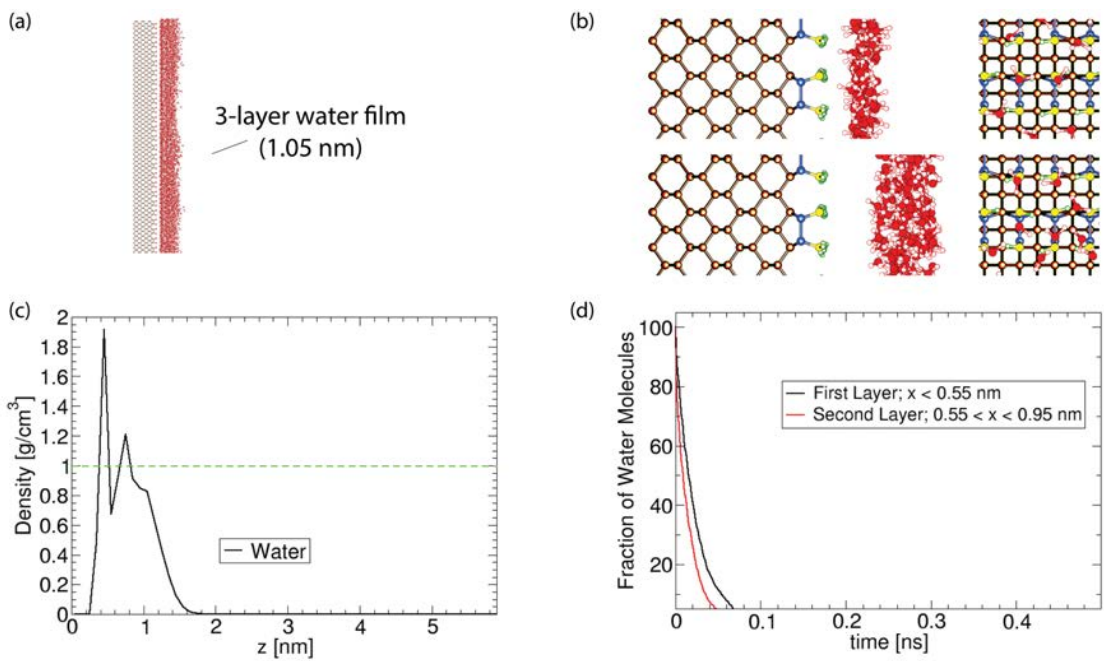

**Fig. S17 | Dynamics of a pure water film.** Snapshot from a MD simulation of a thin water film (approximately 1.0 nm-thick, $N = 4177$ water molecules) on the hydroxylated diamond surface; $T = 300$ K. (b) Density profile of water showing that water molecules arrange in two water layers. (c) Snapshots of the system showing the water molecules in the first ($z < 0.55$ nm; top panel) and second ($z > 0.55$ nm; bollom panel) layers. (d) Fraction of water molecules initially in layers 1 and 2 that remain in the corresponding layer at time $t$. This fraction decays to zero within $< 100$ ps indicating that the water molecules in both layers are able to diffuse over time.

### III.8 *Thin water film*

Partial cavitation or incomplete filling of the nanopore could, in principle, also lead to the slowdown of water molecules. For example, cavitation could lead to the formation of thin films on the confining surfaces, and water molecules in such films could show low or negligible dynamics. To explore this possibility, we also perform MD simulations of a thin water film on the hydroxylated diamond surfaces (Fig. S17a). The water film is ~1 nm-thick and is composed of $N = 4177$ water molecules. After a few picoseconds, water molecules rearrange into two water layers (Fig. S17b); snapshots of water molecules in each layer are shown in Fig. S17c. We find that the water molecules in both layers are able to diffuse. To show this, we include in Fig. S17d the fraction of water molecules in layers $i = 1,2$ at a time $t = 0$ that remain in layer $i$ as function of time. For both layers, we find that the fraction drops to nearly zero for $t < 100$ ps indicating that, indeed, water molecules diffuse over time.